\documentclass[11pt]{aastex631}
\usepackage{amsmath,amsfonts,amssymb}
\usepackage{mathrsfs}
\DeclareMathAlphabet{\mathpzc}{OT1}{pzc}{m}{it}

\usepackage[]{graphicx, epstopdf}
\graphicspath{{Figures/}}

\usepackage{afterpage}

\newcommand{\Ji}{\mathrm{Ji}}

\begin{document}

\title{The Mysterious Lives Of Speckles. I. Residual atmospheric speckle lifetimes in ground-based coronagraphs}

\author{Jared R. Males}
\affiliation{Steward Observatory, University of Arizona, Tucson, 933 N Cherry Ave, Tucson, AZ 85721, USA}

\author[0000-0002-0176-8973]{Michael P. Fitzgerald}
\affiliation{Department of Physics \& Astronomy, University of California, Los Angeles, CA 90095, USA}

\author{Ruslan Belikov}
\affiliation{NASA-Ames Research Center, Moffett Blvd., Moffett Field, CA, USA}

\author{Olivier Guyon}
\affiliation{Astrobiology Center, National Institutes of Natural Sciences, 2-21-1 Osawa, Mitaka, Tokyo, JAPAN}
\affiliation{Steward Observatory, University of Arizona, Tucson, 933 N Cherry Ave, Tucson, AZ 85721, USA}
\affiliation{College of Optical Science, University of Arizona, 1630 E University Blvd, Tucson, AZ 85719, USA}
\affiliation{National Astronomical Observatory of Japan, Subaru Telescope, National Institutes of Natural Sciences, Hilo, HI 96720, USA}

\begin{abstract}
High-contrast imaging observations are fundamentally limited by the spatially and temporally correlated noise source called speckles.  Suppression of speckle noise is the key goal of wavefront control and adaptive optics (AO), coronagraphy, and a host of post-processing techniques.  Speckles average at a rate set by the statistical speckle lifetime, and speckle-limited integration time in long exposures is directly proportional to this lifetime.  As progress continues in post-coronagraph wavefront control, residual atmospheric speckles will become the limiting noise source in high-contrast imaging, so a complete understanding of their statistical behavior is crucial to optimizing high-contrast imaging instruments. Here we present a novel power spectral density (PSD) method for calculating the lifetime, and develop a semi-analytic method for predicting intensity PSDs behind a coronagraph.  Considering a frozen-flow turbulence model, we analyze the residual atmosphere speckle lifetimes in a MagAO-X-like AO system as well as 25--39 m giant segmented mirror telescope (GSMT) scale systems.  We find that standard AO control shortens atmospheric speckle lifetime from $\sim$130 ms to $\sim$50 ms, and predictive control will further shorten the lifetime to $\sim$20 ms on 6.5 m MagAO-X. We find that speckle lifetimes vary with diameter, wind speed, seeing, and location within the AO control region.  On bright stars lifetimes remain within a rough range of $\sim$20 ms to $\sim$100 ms.  Due to control system dynamics there are no simple scaling laws which apply across a wide range of system characteristics.  Finally, we use these results to argue that telemetry-based post-processing should enable ground-based telescopes to achieve the photon-noise limit in high-contrast imaging.
\end{abstract}
 
\section{Introduction}
Characterization of extrasolar planets with resolved imaging is limited by the halo of starlight scattered onto the image plane by optical aberrations.  In addition to the photon noise caused by this halo, the scattered light is coherent and produces ``speckles,'' which can be thought of as copies of the instrument point spread function (PSF) \citep{1995PASP..107..386M}.  This adds structured noise, with both spatial and temporal correlations, which is the limiting noise source in long exposure coronagraphic observations.    Both in space and on the ground, imperfections within the telescope and instrument optics cause speckles.   For ground-based telescopes, the aberrations also include those imposed by atmospheric turbulence.

For some time, long-lived ``quasi-static'' speckles have been recognized as the limiting noise source in contemporary high contrast imaging observations \citep[e.g.][]{2007ApJ...654..633H}.  Given their origin within the instrument, and comparatively long lifetimes \citep{2012A&A...541A.136M}, such speckles should be suppressible using wavefront sensing and control (WFS\&C) strategies.  Recent progress in applying ``dark hole'' contrast optimization with focal plane WFS \citep{2020A&A...638A.117P} and various post-coronagraph focal-plane \citep[e.g.][]{2010A&A...509A..31G} and low-order  WFS\&C strategies \citep[e.g.][]{2017PASP..129i5002S} show great promise in significantly reducing the impact of internal speckles.  Once active quasi-static speckle control becomes routine, the fundamental sensitivity limit in ground-based high-contrast imaging will be set by the Earth's atmosphere.

Speckles also arise from atmospheric turbulence.  The index of refraction of air depends on temperature, and turbulence causes the air temperature to stochastically vary across the telescope aperture.  As the stellar wavefront traverses these regions of different index of refraction, parts of the wavefront travel different optical path lengths.  Due mainly to wind-driven bulk flow across the telescope aperture, the atmospheric temperature structure is constantly changing.  The technique of adaptive optics (AO) is used to correct the resulting time-variable optical aberrations in real time.  Even with so-called ``extreme'' AO systems \citep[ExAO, ][]{2018ARAA..56..315G} this correction is imperfect.  The AO-corrected wavefront, though significantly improved, still causes a halo of speckles which will be significantly brighter than the instrumental speckles in a high performance coronagraph system employing the WFS\&C strategies discussed above.

In order to estimate the capabilities of ground-based high-contrast imaging, we must understand the spatial and temporal evolution of the residual atmospheric speckles behind an ExAO coronagraph.  Speckle formation in the spatial domain, as structures within the coronagraphic PSF, is well understood \citep{2001ApJ...558L..71B,2003ApJ...596..702P,2007ApJ...669..642S}.  In the time domain, we can imagine two ideal regimes for these structures.  In the first regime, the speckles are perfectly static.  Were this the case, it would be a simple matter of measuring the PSF structure and then subtracting it from all images.  While this would still leave the photon noise from the scattered light, we could perfectly remove the spatially correlated noise.  In the second idealized regime, the speckles are a perfect temporally white noise source: every image taken has a different speckle structure.  Similar to photon noise, high-contrast imaging would now be a simple matter of taking enough images to average the speckles to a smooth halo.

In reality, the temporal behavior of atmospheric speckles falls between these regimes. These speckles vary on short time-scales so they are not removed by standard PSF subtraction, but exhibit non-negligible temporal correlations which cause them to average much more slowly than photon noise.  The statistical lifetime of atmospheric speckles, that is the time-scale which controls the averaging of speckle noise, is the quantity that sets the fundamental sensitivity limit of ground-based high-contrast imaging.

\subsection{Atmospheric Coherence}
There have been several definitions of the atmospheric speckle lifetime presented in the literature.  Early treatments of speckle dynamics were motivated by speckle interferometry, which relies on exposures short enough to ``freeze the atmosphere.''  The optimum exposure time is therefore dependent on the correlation time of the intensity in the focal plane.  A common definition for this correlation time was based on a $1/e$ reduction in the autocorrelation of the intensity, which was experimentally found to be on the order of a few ms \citep{1978ApOpt..17.3779S,1990JMOp...37.1247D}.    \citet{1982JOpt...13..263R} derived the following expression for the autocorrelation e-folding time in a multi-layer Kolmogorov atmosphere:
\begin{equation}
\tau_{boil} = 0.36 \frac{r_0}{\sqrt{\sigma^2_\mathcal{V}}}
\label{eqn:e-fold}
\end{equation}
called there the ``speckle boiling time,'' where $r_0$ is Fried's parameter and $\sigma^2_\mathcal{V}$ is the variance with respect to height of the layer wind velocities $\mathcal{V}$  above the telescope.  At good astronomical sites, values of $\tau_\text{boil}$ should range from $\sim$5 to $\sim$20 ms. \citet{1999PASP..111..587R} employed Equation (\ref{eqn:e-fold}) for the speckle lifetime and argued that atmospheric speckles dominate photon noise by factors of 100 or more in high-contrast imaging.

\subsection{Variance of the Mean}
The autocorrelation e-folding time is a somewhat arbitrary choice.  If we instead consider the evolution of the variance in the estimate of the mean given uncorrelated trials, we can define the speckle lifetime according to the familiar ``root-N'' reduction in noise:
\begin{equation}
\sigma_{mean}^2 = \frac{\sigma_o^2}{t/\tau}
\label{eqn:varmean_def_intro}
\end{equation}
where $\tau$ is the noise lifetime, $\sigma_o^2$ is the variance of a single measurement, $\sigma_{mean}^2$ is the variance in the estimate of the mean after time $t$.  This simply states that the variance in a measurement of the mean improves with the number of uncorrelated trials $t/\tau$.  Defined this way, $\tau$ is given by the integral of the autocorrelation \citep{2006ApJ...637..541F}, and in this expression determines the number of uncorrelated realizations of the speckle pattern included in the mean.  Note that this assumes $t \gg \tau$.  \citet{1986JOSAA...3.1001A} employed this definition, and argued for longer speckle lifetimes, replacing $0.36$ by $~$$1.14$ in Equation (\ref{eqn:e-fold}) based on arguments similar to \citet{1982JOpt...13..263R}.  Using AO-corrected images, \citet{2006ApJ...637..541F} empirically found significantly longer values of $\tau$, giving an overall median of 175 ms at Lick observatory.  

\citet{2005SPIE.5903..170M} considered this time averaging of residual atmospheric speckles behind a coronagraph, directly analyzing the reduction in variance as a function of exposure time.  They argued that the speckle lifetime should not be affected by the properties of the AO system nor by the seeing parameter $r_0$, and instead should be determined by the wind-crossing time of a telescope with diameter $D$.  Based on a simulation, \citet{2005SPIE.5903..170M} found that 
\begin{equation}
\tau \approx 0.3 \frac{D}{\mathcal{V}}
\end{equation}
for use in Equation (\ref{eqn:varmean_def_intro})\footnote{\citet{2005SPIE.5903..170M} reported a speckle decorrelation time which would be a factor of two longer than the statistical lifetime}.  This result is quite different from the previous predictions, but coarsely in agreement with the experiments by \citet{2006ApJ...637..541F} (it is challenging to make more definitive experimental comparisons due to the uncertainty in atmospheric parameters).

\citet{2006OExpr..14.7499P} analyzed speckle noise correlations under closed-loop control, in similar fashion to what follows.  Their study found that the wavefront sensor noise and input turbulence contributed differently to speckle lifetime.  They also proposed that the control-loop would impact the lifetime, through the optimization of the control law gain, and that considering long-exposure sensitivity (i.e. through Equation \ref{eqn:varmean_def_intro}) may lead to different optimizations than by considering raw contrast alone. 

More recent experimental studies have used on-sky data with modern ExAO systems.  Using focal plane data from SPHERE, \citet{2016SPIE.9909E..4ZM} found evidence for speckle decorrelation on several timescales, but were unable to identify an atmospheric component given the various instrument-induced timescales and the relatively low cadence used. In a study using SCExAO at much higher frame rates, \citet{2018PASP..130j4502G} analyzed the evolution of the intensity using the variance of difference images (a form of autocorrelation).  Here short timescale (a few ms) speckle evolution was attributed to the atmosphere, and the authors argued that the AO system modulated the speckles on longer timescales.  \citet{2017JATIS...3b5001S} present a study of speckle lifetimes at the LBT, where they found short-lived speckles attributed to the atmosphere and show some evidence for a difference in lifetime inside and outside the AO control radius.

\subsection{Goals of This Study}
Our goal in this study is to develop a self-consistent description of the time-domain behavior of speckle noise in AO-fed coronagraphic images from as close to first principles as possible.  Our motivation is to understand the true limits of ground-based high-contrast imaging and so we focus on the statistical lifetimes of atmospheric speckles in coronagraphic focal planes.   This requires a traversal of speckle statistics, signal processing, wavefront propagation, and control theory. To accomplish our goals we will work through the following tasks, organized by section:
\begin{itemize}
\item Section \ref{sec:noise}: Review speckle statistics to understand how atmospheric speckles dramatically reduce sensitivity.
\item Section \ref{sec:psd_lifetime}: Develop an analytic framework for analyzing power spectral densities (PSDs), which allows us to derive an expression for the statistical (mean-averaging) lifetime from a PSD.
\item Section \ref{sec:intensity}: Show how to calculate the intensity PSD in the coronagraphic focal plane behind a closed-loop AO system with a semi-analytic model.
\item Section \ref{sec:results}: Employ an end-to-end AO system simulation to (a) validate the statistical lifetime calculations and (b) assess the performance of the semi-analytic model.
\item Section \ref{sec:trends}: Use the analytic framework to analyze speckle lifetime over a range of telescope diameters and atmospheric parameters.
\item Section \ref{sec:reconstruction}: Develop a method to analyze post-processing in the time-domain and discuss its implications for the limits of ground-based high-contrast imaging sensitivity.
\item Section \ref{sec:discussion}: Discuss various implications of these results.
\end{itemize}
The key conclusions can be gleaned from Sections \ref{sec:noise}, \ref{sec:trends}, and \ref{sec:discussion}. Sections \ref{sec:psd_lifetime} through \ref{sec:results} and Section \ref{sec:reconstruction} contain the technical details.  

\section{Noise Dynamics in Coronagraphic Focal Planes}
\label{sec:noise}
A key metric for sensitivity in a high-contrast imaging observation is the signal-to-noise ratio, S/N.  The intensity of a planet, which has planet to star contrast of $C_p$, is 
\begin{equation}
I_p(t) = I_{*}(t) C_p
\label{eqn:signal}
\end{equation}
where $I_{*}$ is the peak intensity of the star's PSF without a coronagraph, which is a function of time due to Strehl ratio variation.  Here we will assume units of photon/sec/$(\lambda/D)^2$ for the intensity, where $\lambda$ is the observation wavelength and $D$ is the telescope diameter.  $\lambda/D$ is the fundamental spatial scale of diffraction limited imaging and coronagraphy, and we assume that photometry is conducted in pixels of size 1 $\lambda/D$.

The dominant noise source in such an observation is the halo of scattered starlight underlying the planet signal.  At the planet position $\vec{r}_p$ this has intensity $I_h(\vec{r}_p,t)$. Following \citet{2007ApJ...669..642S} we initially consider three contributions to the halo intensity:
\begin{equation}
I_h(\vec{r}_p,t) = I_{*}(t)\left[\mathcal{I}_c(\vec{r}_p,t) + \mathcal{I}_{as}(\vec{r}_p,t) + \mathcal{I}_{qs}(\vec{r}_p,t)\right]
\end{equation}
$\mathcal{I}_c$ is the fractional contribution from residual diffraction from the coronagraph, as well as purely static aberrations.  $\mathcal{I}_{as}$ is the fractional contribution from residual atmospheric speckles, and $\mathcal{I}_{qs}$ is the fractional contribution from instrumental speckles, which are usually called quasi-static speckles.  
From this we can state the instantaneous raw halo contrast
\begin{equation}
C_h(\vec{r}_p,t) = \frac{I_h(\vec{r}_p,t)}{I_*(t)} = \mathcal{I}_c(\vec{r}_p,t) + \mathcal{I}_{as}(\vec{r}_p,t) + \mathcal{I}_{qs}(\vec{r}_p,t).
\end{equation}

The halo contributes noise in two distinct ways: photon noise and correlated speckle noise. The total halo variance, considering both speckles and and photon noise, under the planet signal is \citep{2007ApJ...669..642S} 
\begin{equation}
\sigma^2_{h}  = I_* \Delta t [\underbrace{\mathcal{I}_c + \mathcal{I}_{as} + \mathcal{I}_{qs}}_{\mbox{photon noise}} + I_* [ \underbrace{\tau_{as}\left(\mathcal{I}_{as}^2 + 2[\mathcal{I}_c\mathcal{I}_{as} + \mathcal{I}_{as}\mathcal{I}_{qs}]  \right)}_{\mbox{atm. speckles}} +  \underbrace{\tau_{qs}\left( \mathcal{I}_{qs}^2 + 2 \mathcal{I}_c \mathcal{I}_{qs}\right)}_{\mbox{QS speckles}}]]
\label{eqn:variance}
\end{equation}
where $\tau_{as}$ is the lifetime of the residual atmospheric speckles, and $\tau_{qs}$ is the lifetime of the instrumental quasi-static speckles.  For simplicity we have suppressed the $\vec{r}_p$ dependence.   

We have likewise dropped the $t$ dependence, but we note that this is somewhat subtle.  We could assume that $\Delta t$ is sufficiently short that the each of the $\mathcal{I}$ terms can be treated as constant, and then state the long exposure variance as a sum of the variance in these short exposures.  Alternatively, we can treat the $\mathcal{I}$ terms in Equation (\ref{eqn:variance}) as mean values, and then $\sigma^2_{h}$ is the mean of the variance.  For large total exposure time these will converge.

Our present focus is on the atmospheric residual $\mathcal{I}_{as}$.  From here on we will assume a perfect coronagraph and instrument, such that $\mathcal{I}_c = 0$ and $\mathcal{I}_{qs}$ = 0.  These (optimistic) conditions eliminate the speckle pinning \citep{2001ApJ...558L..71B} terms in Equation (\ref{eqn:variance}).  The halo variance is then due solely to atmospheric speckles:
\begin{equation}
\sigma^2_{h}  = I_* \Delta t \left(\mathcal{I}_{as} + I_* \tau_{as} \mathcal{I}_{as}^2 \right)
\label{eqn:variance_as}
\end{equation}

The noise is the square-root of the variance, so using Equations (\ref{eqn:signal}) and  (\ref{eqn:variance_as}) the peak signal-to-noise ratio of the planet detection will be
\begin{equation}
\mbox{S/N} = \frac{C_p \sqrt{I_* \Delta t}}{ \sqrt{\mathcal{I}_{as} (1 + I_*\mathcal{I}_{as} \tau_{as})}} 
\end{equation}
For argument's sake, assume a star having $I_* = 10^8$ photon/sec/$(\lambda/D)^2$ in the peak (roughly an 8th mag star on a 10 m telescope in the near-IR), and $\mathcal{I}_{as} = 1\times10^{-4}$ (see \citealt{2018JATIS...4a9001M}).  Next we (for now) assume the \citet{2005SPIE.5903..170M} crossing-time speckle lifetime for a 10-m telescope with 10-m/s wind, giving $\tau_{as}$ = 0.3 s.  We now have $I_*\mathcal{I}_{as} \tau_{as} = 3000$.  We see that under such conditions  $I_*\mathcal{I}_{as} \tau_{as} >> 1$, which would hold even for significantly smaller $\tau_{as}$.  This leads us to
\begin{equation}
\mbox{S/N} \approx \frac{C_p \sqrt{\Delta t}}{ \mathcal{I}_{as} \sqrt{\tau_{as}}}. 
\end{equation}
Such an observation is therefore \emph{speckle-noise limited}, and the S/N improves with the square-root of the number of realizations of the speckle pattern as $\sqrt{\Delta t/ \tau_{as}}$.  The ratio of the time to reach a given S/N when speckle limited, $\Delta t_{sl}$, to the time needed when photon noise limited, $\Delta t_{pn}$ is
\begin{equation}
\frac{\Delta t_{sl}}{\Delta t_{pn}} = I_*\mathcal{I}_{as} \tau_{as} \gg 1.
\label{eqn:relexptime}
\end{equation}

Understanding the speckle lifetime is therefore crucial to understanding the limits of high-contrast imaging.  To that end, we next show how to calculate the statistical correlation time in a stochastic process from its PSD, and then show how to apply it to post-coronagraph speckles.

\section{Speckle Lifetime From A PSD}
\label{sec:psd_lifetime}
Starting with the intensity time series $I(t)$ measured at the location of a planet, the temporal PSD $\mathcal{T}(f)$ is defined as the ensemble average of the modulus-squared of its Fourier transform:
\begin{equation}
\mathcal{T}(f) = \left\langle \left| \mathcal{F}_t \left\{ I(t) \right\} \right|^2 \right\rangle.
\label{eqn:psd_def}
\end{equation}
where $\mathcal{F}_t\{\cdot\}$ denotes the Fourier transform in one dimension (the time-domain) and $f$ is temporal frequency.  We will restrict ourselves to one-sided PSDs, that is $f \ge 0$.  The PSD is normalized such that the variance of measurements from a process governed by the PSD is
\begin{equation}
\sigma_o^2 = \int_{0}^{f_s/2} \mathcal{T}(f) df.
\label{eqn:process_var}
\end{equation}
where $f_s$ is the sampling frequency defined by the sampling rate $\Delta t$ according to $f_s = 1/\Delta t$.

We assume that $I(t)$ is a wide-sense stationary (WSS) process.  This means that the mean is constant and the autocorrelation does not depend on time.   In this context, WSS essentially implies that the atmosphere parameters, namely $r_0$ and the layer $\vec{\mathcal{V}}$, are constant.  This is of course not strictly true, but can be expected to be approximately true over short periods of time, by which we mean at least long enough to estimate a PSD.  WSS also implies that AO control is stable, and we expect the control system to damp changes, keeping the output speckle formation process within a more or less well-defined range.

With the WSS assumption, the Wiener-Khinchin theorem states that the PSD is the Fourier transform of the autocorrelation. The PSD then describes how the measurements derived from the process are correlated. \citet{1976JOSA...66..207N} exploited this concept in two dimensions to derive the coefficient covariances in the Zernike polynomial expansion of Kolmogorov turbulence phase screens.  Here we adapt that analysis to arbitrary one dimensional PSDs using the Legendre polynomials.  

\begin{figure}[h]
    \centering
    \includegraphics[width=3.25in]{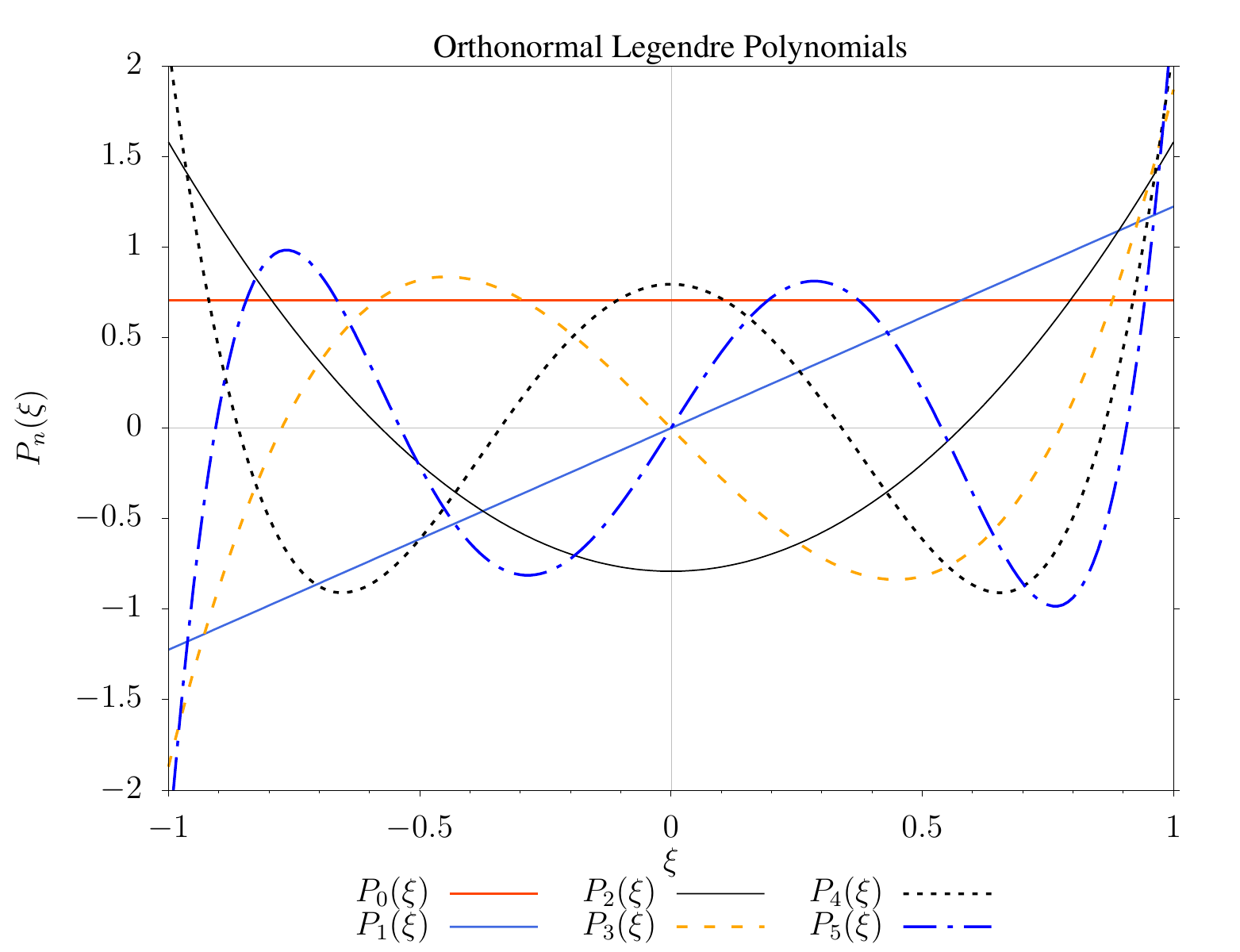}
    \caption{The orthonormal Legendre polynomials, which are used as a basis for analyzing the intensity time-series.  We show the first six. \label{fig:onlegpol}}
\end{figure}

In principle, any basis could be used.  Our main goal is analyze the statistics of the mean, which corresponds to the first Legendre polynomial.  Similar to the Zernike polynomials, this basis has an intuitive progression through slope, quadratic, cubic, etc, modes (see Figure \ref{fig:onlegpol}). We later utilize these one dimensional modes to analyze post-processing as a time-domain filter.  For that analysis it will be convenient to use a normalized form of Legendre polynomials, defined as
\begin{equation}
P_{\eta}(\xi) = \sqrt{\frac{2\eta+1}{2}}\, \mathscr{P}_\eta(\xi),
\label{eqn:basis}
\end{equation}
where $\mathscr{P}_\eta(\xi)$ are the Legendre polynomials defined in the usual manner \citep[cf.][]{Olver:2010:NHMF} on the domain $-1 \le \xi \le 1$. The first six of these functions are shown in Figure \ref{fig:onlegpol}.  With our normalization we have the orthogonality condition
\begin{equation}
\int_{-1}^{1}P_\eta(\xi)P_{\eta'}(\xi) d\xi = \delta_{\eta\eta'}
\label{eqn:orthonormal}
\end{equation}
where $\delta_{\eta\eta'}$ is the Kronecker delta. 

Now we obtain a sample of $I(t)$ of length $T$ consisting of measurements uniformly spaced by $\Delta t$.  For simplicity in notation we define the time coordinate such that $-T/2 \le t \le T/2$, with no loss of generality.  Next, we express this time-series as a linear expansion using the normalized Legendre polynomials.  Changing variables with
\begin{equation}
\xi = \frac{2}{T}t
\end{equation}
we have
\begin{equation}
I(t) = I\left(\frac{T}{2}\xi\right) = \sum_\eta a_\eta P_\eta(\xi)
\label{eqn:basisexpansion}
\end{equation}
with coefficients
\begin{equation}
a_\eta = \int_{-1}^{1} I\left(\frac{T}{2}\xi\right)P_\eta(\xi) d\xi.
\label{eqn:expansecoeff}
\end{equation}
The coefficients $a_\eta$ are variable quantities, meaning that they will change with time as the speckle intensity varies.  By analyzing the ensemble statistics of the $a_\eta$, we learn about the power in specific Legendre ``modes'' over the time $T$ just as we do with the analogous Zernike modes in a wavefront over an aperture of diameter $D$ \citep{1976JOSA...66..207N}.

The first orthonormal Legendre polynomial is $P_0(\xi) = 1/\sqrt{2}$, which means that $a_0/\sqrt{2}$ is the mean value of $I(t)$ over a time-period of length $T$.  Thus the variance of the mean is given by
\begin{equation}
\sigma_\mathrm{mean}^2 = \frac{\left\langle \left|a_0\right|^2 \right\rangle}{2}
\label{eqn:varmean_def}
\end{equation}
where $\langle \cdot \rangle$ denotes the ensemble average.   

Next we use the techniques developed in \citet{1976JOSA...66..207N} to calculate $\left\langle \left|a_0\right|^2 \right\rangle$. The step-by-step derivation is given in Appendix \ref{app:tau_psd}, and here we give the result that:
\begin{equation}
\sigma_\mathrm{mean}^2 = \frac{\left\langle \left|a_0\right|^2 \right\rangle}{2} = \frac{1}{2T}  \int_0^{\infty} \frac{ J_{\frac{1}{2}}^2(2\pi \kappa)}{\kappa} \mathcal{T}\left( \frac{2}{T} \kappa \right) d\kappa
\label{eqn:varmean}
\end{equation}
where $J_n$ denotes the cylindrical Bessel functions of the first kind.  This expression gives the variance of the mean in a process governed by $\mathcal{T}(f)$ sampled in chunks of length $T$.

Now consider the case of a white noise process, which has a constant PSD
\begin{equation}
\mathcal{T}(f) = \beta.
\end{equation}
Evaluating Equation (\ref{eqn:varmean}), the variance of the mean after a time $T$ is 
\begin{equation}
\sigma_\mathrm{mean}^2 = \frac{\beta}{2T}.
\label{eqn:vm_wn}
\end{equation}
Now from Equation (\ref{eqn:process_var}) the variance of a single measurement is
\begin{equation}
\sigma_o^2  = \frac{\beta}{2\Delta t}.
\end{equation}
Rearranging and substituting for $\beta$ in Equation (\ref{eqn:vm_wn}) yields
\begin{equation}
\sigma_\mathrm{mean}^2 =  \frac{\sigma_o^2 \Delta t}{T} = \frac{\sigma_o^2}{N}
\end{equation}
where $N = T/\Delta t$ is the number of measurements.  This is exactly the result we expect for the variance of the mean in a white noise process with variance $\sigma^2_o$, giving the familiar ``root-N'' reduction in noise.

To generalize this result we define $\tau$ as the correlation length of the process over the period $T$.  Then for an arbitrary PSD
\begin{equation}
\tau = \lim_{T\to\infty}\frac{ \displaystyle\int_0^{\infty} \frac{ J_{\frac{1}{2}}^2(2\pi \kappa)}{\kappa} \mathcal{T}\left( \frac{2}{T} \kappa \right) d\kappa}{ 2 \displaystyle\int_{0}^{f_s/2} \mathcal{T}(f) df}.
\label{eqn:psd_lifetime}
\end{equation}
The variance of the mean then decreases with time according to
\begin{equation}
\sigma_\mathrm{mean}^2 =  \frac{\sigma_o^2}{t/\tau} \tag{\ref{eqn:varmean_def_intro} restated}
\end{equation}
as long as $t \gg \tau$.

Equation (\ref{eqn:psd_lifetime}) shows that the statistical lifetime is mainly set by the lowest frequency components of the PSD.  The limit in the numerator means that $f = \frac{2}{T}\kappa \rightarrow 0$, so we can approximate the lifetime as $\tau \approx \mathcal{T}(0)/(2\sigma_o^2)$.  That is, the lifetime is set by the proportion of power in the 0-frequency component of the PSD.  As we will show, both real-time AO control and post-processing act to reduce power in the low frequencies of the PSD, so we therefore expect these processes to reduce the speckle lifetime.

Similar conclusions have been found through other approaches.  The statistical lifetime can be found from the integral of the autocovariance \citep{1986JOSAA...3.1001A}, which gives the 0-frequency component of the PSD (this is essentially how the derivation in Appendix \ref{app:tau_psd} proceeds).  The lifetimes so derived have the same time-averaging interpretation we use here \citep{2006ApJ...637..541F}.  \citet{2006OExpr..14.7499P} used the moving average filter to analyze the time evolution of variance in atmospheric speckles \citep[see also][]{2005SPIE.5903..170M}.  This likewise selects the lowest frequency components as integration time grows, and so produces equivalent results.

\section{Speckle Intensity Temporal PSDs}
\label{sec:intensity}
\subsection{Intensity as a Function of Time}
Here we very briefly review the derivation of the intensity time-series from \citet{2018JATIS...4a9001M}, with the goal to introduce the notation needed for this analysis.  See that paper for a detailed treatment.

We use a real-valued Fourier basis
\begin{equation}
M_{mn}^p(\vec{q}) = \cos(2\pi \vec{k}_{mn} \cdot \vec{q}) + p  \sin(2\pi \vec{k}_{mn} \cdot \vec{q})
\label{eqn:modified_fourier}
\end{equation}
where $p=\pm 1$, $\vec{q}$ is the position vector in the pupil plane and the vector spatial-frequency is 
\begin{equation}
\textstyle\vec{k}_{mn} = \textstyle\frac{m}{D} \hat{\mathpzc{u}} + \textstyle\frac{n}{D} \hat{\mathpzc{v}}
\label{eqn:kq_def}
\end{equation}
where $m$ and $n$ are integer indices\footnote{Note that this $n$ is different from that used in the Legendre indices.  The difference will be clear from the context.}, $D$ is the aperture diameter, and $(\hat{\mathpzc{u}},\hat{\mathpzc{v}})$ are unit vectors.  The modal basis described by Equation (\ref{eqn:modified_fourier}) is normalized, and has the advantage that on the unobscured circular aperture both $p$ modes at a given $mn$ have the same symmetric spatial and temporal PSD.  This is not true for pure sines and cosines.  We note that this basis is the one derived for the unobscured circular aperture in a 2D Lomb-Scargle analysis \citep{2020arXiv200110200S}, and, as such, phase-shifted Fourier bases with the same properties exist for more complicated apertures.  In this modal basis, the phase in the pupil plane is
\begin{equation}
\Phi(\vec{q},t) = \frac{2\pi}{\lambda} \sum\limits_{mn} \left[ h_{mn}^{+}(t) M_{mn}^{+}(\vec{q}) + h_{mn}^{-}(t) M_{mn}^{-}(\vec{q})\right]
\label{eqn:phi_expansion}
\end{equation}
where $h_{mn}^{+}$ and $h_{mn}^{-}$ are the real-valued coefficients of the modes in the same units as wavelength, and the phase $\Phi$ has units of radians.

\citet{2018JATIS...4a9001M} then showed how the residual intensity due to the phase $\Phi$ behind a perfect coronagraph is described by
\begin{equation}
I_\Phi(\vec{r},t) \approx \left(\frac{2\pi}{\lambda}\right)^2 2 \sum_{mn} \left\{  \left[ (h_{mn}^{+}(t))^2 + (h_{mn}^{-}(t))^2 \right] \left[ \Ji^2(\pi D k_{mn}^+) +\Ji^2(\pi D k_{mn}^-)\right] + \mbox{cross-terms} \right\}
\label{eqn:cross_terms}
\end{equation}
at location $\vec{r} = r_x \hat{x} + r_y\hat{y}$ where  $\hat{x}$ and  $\hat{y}$ are the focal plane coordinate unit vectors. $\Ji$ is the Jinc function
\begin{equation}
\Ji(x) = \frac{\mathrm{J}_1(x)}{x}.
\end{equation}
and the modified spatial frequencies are given by
\begin{eqnarray}
k_{mn}^+ &=& \sqrt{\left(\frac{r_x}{\lambda} + \frac{m}{D}\right) + \left(\frac{r_y}{\lambda} + \frac{n}{D}\right)} \\ 
k_{mn}^- &=& \sqrt{\left(\frac{r_x}{\lambda} - \frac{m}{D}\right) + \left(\frac{r_y}{\lambda} - \frac{n}{D}\right)}. \nonumber
\end{eqnarray}
Equation \ref{eqn:cross_terms} is equivalent to a convolution with the PSF, adding contributions from nearby speckles.  The cross-terms are a series of terms including products of $h_{mn}^+h_{mn}^-$ and $h_{mn}^ph_{m'n'}^{p'}$ ($mn \neq m'n'$) with $\Ji(x)\Ji(x')$.  As long as the aperture truncation is accounted for when the long-exposure statistics of $h$ are calculated, the sum can be dropped and these small cross-terms vanish.  With this caveat, the long exposure intensity is 
\begin{equation}
\left< I_\Phi(\vec{r}_{mn}) \right> =  \left(\frac{2\pi}{\lambda}\right)^2\left< |h_{mn}|^2 \right> .
\label{eqn:contrast_h2}
\end{equation}
where the discretized position 
\begin{equation}
\vec{r}_{mn} = \vec{k}_{mn} \lambda
\end{equation}
describes the relationship between the spatial-frequency in the pupil plane and projected-angular position in the focal plane.

\subsection{The Intensity PSD}

The quantity we seek is the temporal PSD of the intensity time-series described by Equation (\ref{eqn:cross_terms}).  The PSD $\mathcal{T}_{I_{mn}}(f)$ is defined as in Equation (\ref{eqn:psd_def}).

The average Fourier transform of a time-series formed from the products of uncorrelated zero-mean processes will be 0, so we can employ the same logic as above and write
\begin{equation}
\mathcal{T}_{I_{mn}}(f) \approx \left(\frac{2\pi}{\lambda}\right)^2  \left\langle \left| \mathcal{F}_t\left\{   (h_{mn}^{+})^2 + (h_{mn}^{-})^2  \right\} \right|^2 \right\rangle  .
\label{eqn:psd_h2}
\end{equation}
We therefore directly relate the temporal PSD of $I_\phi$ to the temporal PSD of $(h_{mn}^{+})^2 + (h_{mn}^{-})^2$. This can be estimated given the PSDs for $h_{mn}^{+}$ and $h_{mn}^{-}$; we review how these can be calculated next.

\subsection{AO Corrected Post-Coronagraph PSDs}

We express the PSDs of the $h_{mn}^p$ in closed loop ($\mathcal{T}_{\mathrm{cl},mn}(f;g)$) as functions of the open-loop (OL) PSD ($\mathcal{T}_{\mathrm{ol},mn}(f)$), the error transfer function (ETF), the measurement noise PSD ($\mathcal{T}_{\mathrm{ph},mn}(f)$), and the noise transfer function (NTF): 
\begin{equation}
\mathcal{T}_{\mathrm{cl},mn}(f;g) = \mathcal{T}_{\mathrm{ol},mn}(f) \left| \mbox{ETF}_{\mathrm{cl}}(f;g) \right|^2 + \mathcal{T}_{\mathrm{ph},mn}(f) \left| \mbox{NTF}_{\mathrm{cl}}(f;g) \right|^2 
\label{eqn:cl_psd}
\end{equation}
where the control loop gain $g$ is optimized to minimize the residual variance in the closed-loop PSD \citep{1999aoa..book.....R_ch6,2016ApOpt..55..323P}. \citet{2018JATIS...4a9001M} derived OL PSDs of Fourier modes in frozen-flow turbulence.  In simplified form, the wavefront sensor (WFS) noise is given by
\begin{equation}
\sigma_{ph}^2 = \frac{\beta_{p,mn}^2}{F_\gamma \tau_\mathrm{wfs}}
\label{eqn:photon_noise}
\end{equation}
where $\beta_{p,mn}$ is the WFS sensitivity to photon noise \citep{2005ApJ...629..592G}, $F_\gamma$ is the total photon rate (photons/sec) and $\tau_\mathrm{wfs}$ is the WFS integration time.  This gives a white-noise PSD
\begin{equation}
\mathcal{T}_{\mathrm{ph},mn}(f) = 2\frac{\beta_{p,mn}^2}{F_\gamma }.
\end{equation}

We consider ETFs and NTFs for control laws based on general linear filters of the form
\begin{equation}
\widetilde{h} (t_i) = \sum_{j=1}^J a_j \widetilde{h}(t_{i-j}) + g\sum_{l=0}^L b_l \Delta h (t_{i-l})  
\label{eqn:linfilt}
\end{equation}
Here the $\widetilde{h}_i$ are the estimates of the mode coefficient going back in time, and the $\Delta h_i$ are the measurements of the closed-loop residual amplitude of the mode.  The filter combines the $J$ previous estimates and $L$ previous measurements to form the optimum estimate for the current time, which is then applied as a DM command.  

With $J=1$, $a_1 = 1$, $L=0$, and $b_0 = 1$, Equation (\ref{eqn:linfilt}) represents the simple integrator (SI).  \citet{2018JATIS...4a9001M} showed how adding more coefficients, using the well-known Linear Prediction (LP) technique, results in significant performance improvements in terms of post-coronagraph contrast.  Most, if not all, linear controllers can be cast in the form of Equation (\ref{eqn:linfilt}) \citep{2007JOSAA..24.2645P, 2021arXiv210307566H}. 

\subsection{The Relationship Between $h_{mn}^+$ and $h_{mn}^-$ in Frozen Flow Turbulence}

An alternate way to describe the time-evolution of the turbulent wavefront is (see \citet{2005ApJ...629..592G} and Appendix B of \citet{2018JATIS...4a9001M})
\begin{equation}
\Phi_{mn}(\vec{q},t) = \frac{2\pi}{\lambda} h_{mn}^\dagger(t) \cos \left( 2\pi \vec{k}_{mn} \cdot \vec{q} + \phi_{mn}(t) \right)
\end{equation}
These relate to the coefficients of the basis defined by Equation (\ref{eqn:modified_fourier}) as
\begin{eqnarray}
h_{mn}^+ &=& \frac{1}{2} h_{mn}^\dagger(t) \left[ \cos(\phi_{mn}(t)) + \sin(\phi_{mn}(t))\right] \\
h_{mn}^- &=& \frac{1}{2} h_{mn}^\dagger(t) \left[ \cos(\phi_{mn}(t)) - \sin(\phi_{mn}(t))\right] \nonumber
\end{eqnarray}
Under the Taylor frozen flow hypothesis the amplitude $h_{mn}^\dagger$ of each spatial frequency is approximately constant over short periods, which means that $h_{mn}^-(\phi_{mn}(t))$ = $h_{mn}^+(\phi_{mn}(t) + \frac{\pi}{2})$.  This phase shift yields
\begin{equation}
\mathcal{F}_t\left\{h_{mn}^-\right\} = \mathcal{F}_t\left\{h_{mn}^+\right\} e^{i\frac{\pi}{2}}.
\end{equation}
This temporal-phase relationship between the two modes at each spatial frequency must be included in the PSD calculation in order to obtain accurate speckle lifetimes.

\subsection{Spatial Correlations}

In the following recipe for generating PSDs, we will produce time-series of the modal coefficients $h_{mn}^p$.  In addition to the temporal phase relationship, there is a modest spatial correlation between the modes, even under good correction.  This is illustrated in Figures 2 and 3 of \citet{2018JATIS...4a9001M}.  This can be accounted for using the covariance matrix $\pmb{\Sigma}$ of the $h_{mn}^p$, which is calculated as described in that paper.  Given an initially uncorrelated vector $\tilde{h}_{mn}^p$, and the decomposition $\pmb{\Sigma} = \pmb{A}\pmb{A}^T$, the spatially correlated vector of coefficients is found from
\begin{equation}
h_{mn}^p = \pmb{A} \tilde{h}_{mn}^p.
\end{equation}

We note that this adds significant computational cost to the following calculations, for relatively minor benefits.  In practice we have found that it only modestly improves the match between these calculations and simulations at higher temporal frequencies, and has little to no noticeable impact on speckle lifetime predictions.  We include it here for completeness, and consider it optional depending on the goals for employing the model.

\subsection{A Recipe for Intensity PSDs}
\label{sec:recipe}
We are now ready to calculate $\mathcal{T}_{I_{mn}}(f)$.  Unfortunately, we have found no purely analytic way to calculate $\langle \left| \mathcal{F}_t\left\{ h^2 \right\} \right|^2 \rangle$ given only $\langle \left| \mathcal{F}_t\left\{ h \right\} \right|^2 \rangle$.  This is further complicated by the spatial correlations and temporal relationship between modes.  We therefore employ a Monte Carlo procedure:
\begin{enumerate}
\item Generate an OL time-series for each $h_{mn}^p$ for both parities.  Note that $h_{mn}^-$ is needed for the optional spatial correlation step, even though it is discarded eventually. \label{step:olh}
\begin{enumerate}
\item Generate a Gaussian white-noise time-series of the desired length
\item Fourier transform the white-noise time-series
\item Multiply the Fourier transform by the square-root of the OL PSD.  We now have $\mathcal{F}_t\left\{h_{mn}^p\right\}$ \citep[cf.][]{kasdin_falpha}.
\item Apply the inverse Fourier transform to produce $h_{mn}^p(t)$.  If the correlation matrix will be applied in Step \ref{step:corr}, then normalize to unit variance.
\item Repeat for each mode.
\end{enumerate}
\item Optionally, correlate the modal time-series by multiplying the vector of coefficients at each time-step by the decomposition of the covariance matrix. \label{step:corr}
\item For each spatial frequency $mn$, replace $h_{mn}^-(t)$ with the phase shifted $h_{mn}^+(t)$.
\item Generate a WFS measurement noise time-series for each mode, $\eta_{mn}^+$ and $\eta_{mn}^-$, and Fourier transform them.  These are white-noise PSDs, with no phase relationship.
\item Generate the closed-loop Fourier transforms for each mode \label{step:clh}
\begin{eqnarray}
\mathcal{F}_t\{h_{mn,cl}^+\} = \mathcal{F}_t\{h_{mn}^+\} \mathrm{ETF}(f) + \mathcal{F}_t\{\eta_{mn}^+\} \mathrm{NTF}(f) \\
\mathcal{F}_t\{h_{mn,cl}^-\} = \mathcal{F}_t\{h_{mn}^-\} \mathrm{ETF}(f) + \mathcal{F}_t\{\eta_{mn}^-\} \mathrm{NTF}(f) \nonumber
\end{eqnarray}
\item Calculate the closed-loop modal coefficient time-series with the inverse Fourier transform.  We now have $h_{mn,cl}^+(t)$ and $h_{mn,cl}^-(t)$ 
\item Now we can calculate the focal plane intensity as a function of time
\begin{equation}
I_{mn}(t) \approx \frac{2\pi}{\lambda}\left[ \left(h_{mn,cl}^+\right)^2 + \left(h_{mn,cl}^-\right)^2 \right] 
\end{equation}
\item Finally, we calculate $|\mathcal{F}_t\{I_{mn}(t)\}|^2$.  For best results this is done with a window function applied.
\item Now the entire procedure is repeated many times, averaging the result to form the estimate of $\mathcal{T}_{I_{mn}}(f)$.
\end{enumerate}

In the following sections we compare the results of these calculations to end-to-end numerical simulations and analyze the results in terms of speckle lifetimes.



\section{Results}
\label{sec:results}


\subsection{Simulations}
\label{sec:simulations}
We conducted a series of end-to-end numerical simulations of an idealized instrument to validate our calculations, using MagAO-X \citep{2020SPIE11448E..4LM} as an example.  Our goal is not to develop a realistic hardware simulation, rather to ensure that the the calculations produce reasonable results. 

We simulated a circular unobscured aperture of 6.5 m.   An idealized deformable mirror was used, which exactly produced commanded shapes as projections of modes.  The DM was capable of making 2,400 Fourier modes, corresponding to the Nyquist limit of a 48$\times$48 actuator illuminated pupil.  The wavefront sensor (WFS) worked by simply projecting modes onto the phase screen.  The basis set used for control was the modified Fourier basis (Equation (\ref{eqn:modified_fourier})) but orthonormalized with the stabilized Gramm-Schmidt procedure.  This was necessary for the loop to be stable under the LP controller.  Photon noise was added as Gaussian noise uniformly distributed across the WFS image.  The noise was scaled to correspond to the unmodulated pyramid sensor  \cite[$\beta_p = \sqrt{2}$,][]{2005ApJ...629..592G}.

Wavefront propagation was monochromatic using the Fraunhofer approximation, and the science and sensing wavelength were both 800 nm.  The WFS was spatially filtered in the Fourier domain to minimize aliasing.  The perfect coronagraph was simulated \citep{2006A&A...447..397C}.  The wavefront was spatially filtered before the coronagraph as well, to minimize Gibbs ringing on the dark hole edge due to the digitized pupil.

The simulated system operated at 2,000 Hz with a 1.5 frame delay in addition to the 1 frame total sample and hold for WFS integration.  To achieve the 1/2 frame, the WFS used the average turbulence between two time-steps.  Turbulence was generated in 7 layers as fixed oversized phase screens, translated according to the layer wind speed.   Turbulence parameters corresponded to the median LCO model used in \citet{2018JATIS...4a9001M}, where $r_0 = 0.16$ m, $L_0 = 25$ m, and layer averaged wind speed was 18.7 m/s.  The control laws, both SI and LP with 21 coefficients, were optimized on the OL PSDs measured by running the simulation with the control loop off.

Simulations were conducted first for an infinitely bright star, that is without photon noise, to analyze the impact of only the input disturbance dynamics.  Next, an 8th magnitude star was simulated to show the effects of photon noise in the WFS.  For both cases, simulations with both the SI and LP controllers were conducted.  Individual simulation runs covered an elapsed time of 65 seconds, and the final 60 seconds were analyzed to be sure that the control system had stabilized.  Each of these four cases were run 10 times with a new turbulence phase screen, and the results averaged.

\subsection{Raw Contrast}
\begin{figure}
\hspace{-0.3in}
\centering
\includegraphics[width=3.54in]{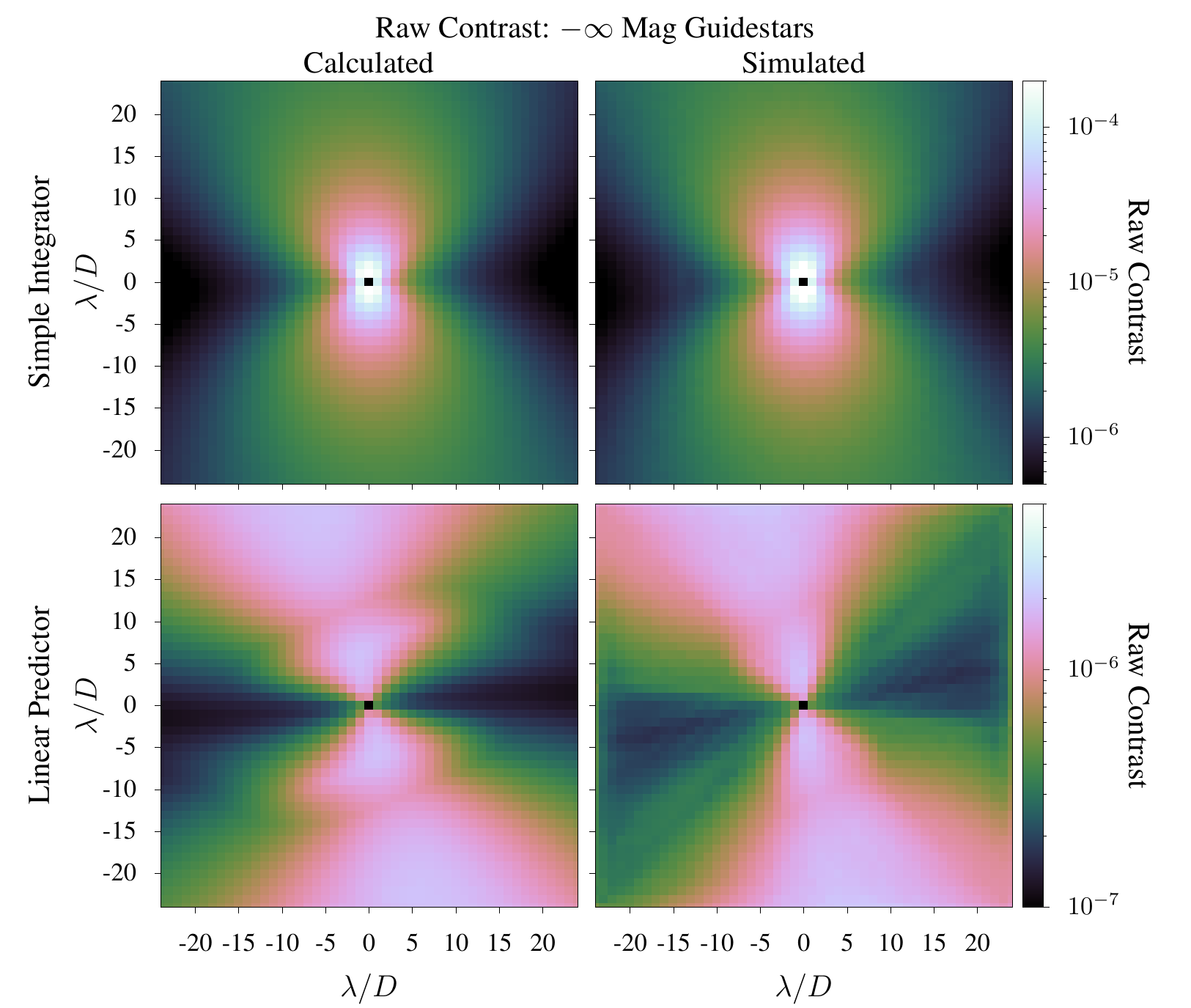}
\includegraphics[width=3.54in]{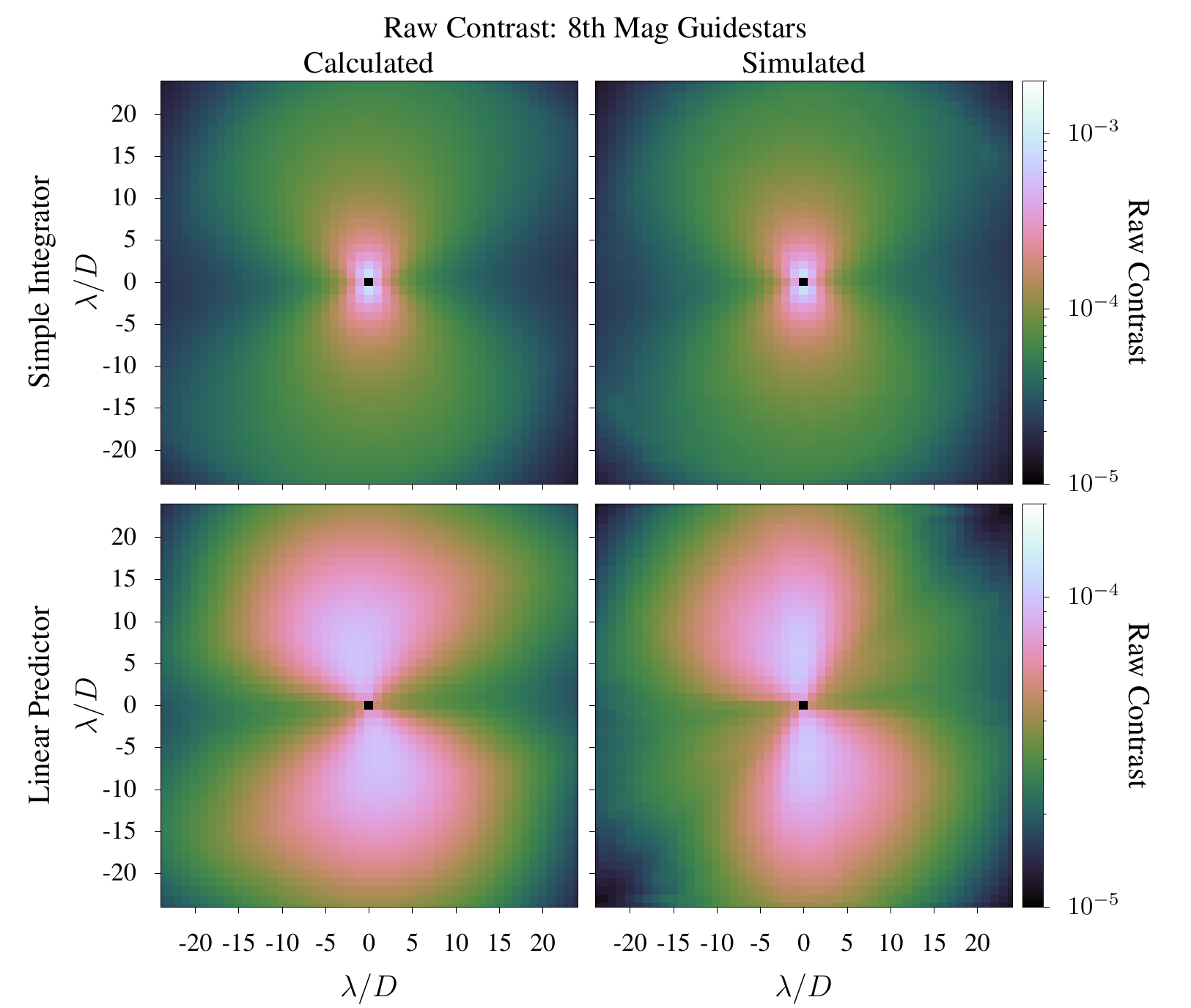}
\caption{Comparison of long-exposure raw contrast from semi-analytic calculations (Section \ref{sec:recipe}) and end-to-end simulation (Section \ref{sec:simulations}) for a 6.5 m telescope. Left: without photon noise, considering the input turbulence only.  Right: with photon noise on an 8th mag star assuming the sensitivity of an unmodulated pyramid WFS. The same colorbar is used for both calculated and simulated within a combination of controller and guidestar, but is adjusted to best show detail for each combination.  The agreement between the calculations and simulations shows that the model is accurate.\label{fig:contrast_2Dcomp}}
\end{figure}

\begin{figure}
\hspace{-0.3in}
\centering
\includegraphics[width=6.5in]{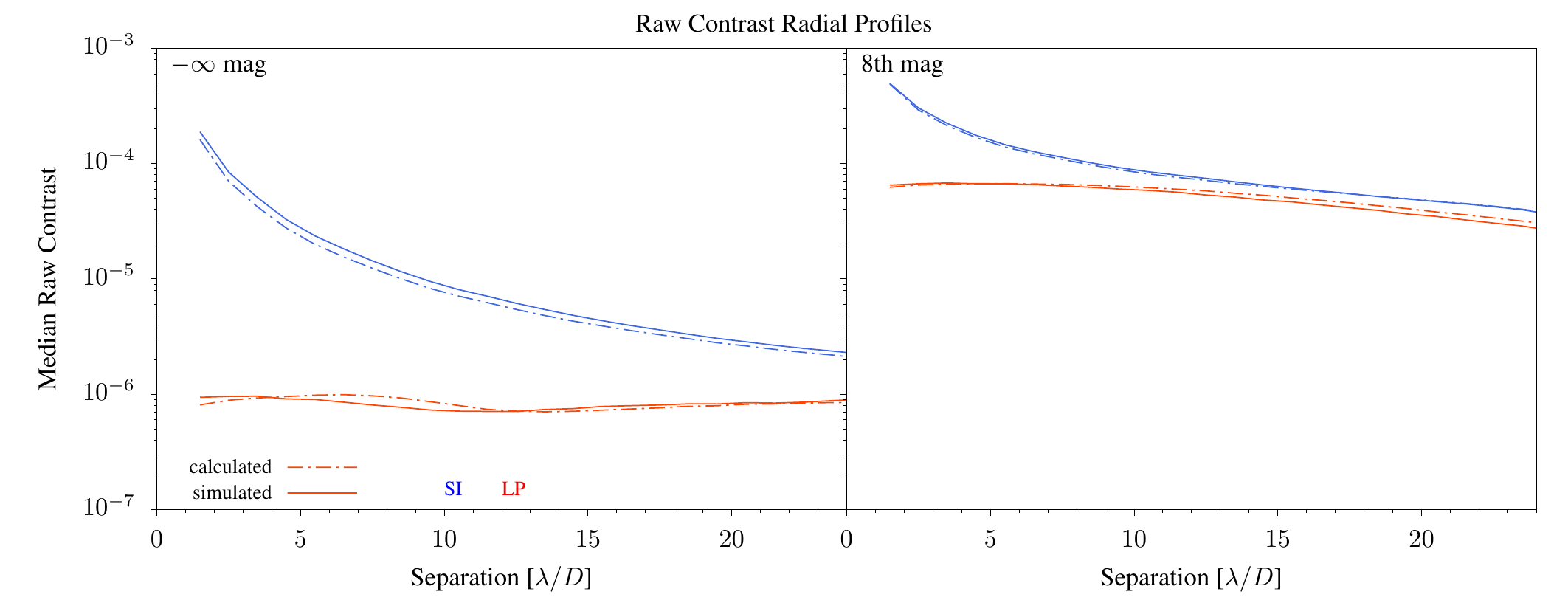}
\caption{Median radial profiles of the contrast maps shown in Figure \ref{fig:contrast_2Dcomp}.  Left: without photon noise, considering the input turbulence only.  Right: with photon noise on an 8th-mag star assuming the sensitivity of an unmodulated pyramid WFS. The agreement between the calculations and simulations shows that the model is accurate. \label{fig:contrast_radprof}}
\end{figure}

Figure \ref{fig:contrast_2Dcomp} shows post-coronagraph contrast maps, comparing the results of calculations using \citet{2018JATIS...4a9001M} to the simulation outputs.  We show results for an infinitely bright ($-\infty$ mag) guide star to show the effects of the input turbulence alone, and for an $8$th-mag guide star.   Figure \ref{fig:contrast_radprof} shows the median radial profile for each of the maps in Figure \ref{fig:contrast_2Dcomp}.  The agreement is very good, though some small differences are visible in the LP (bottom row). These differences, apparent in both the maps and the profile, are due to the LP controller optimization leading to slightly different controller characteristics.  This is due to the difference in the modal basis actually under control in the simulations compared to the model. Another possible source of slight differences is spatial-temporal correlation not fully accounted for in the semi-analytic calculations, which may become apparent at the correction levels under predictive control.

\subsection{Dynamics}
\begin{figure}
\hspace{-0.3in}
\centering
\includegraphics[width=6.5in]{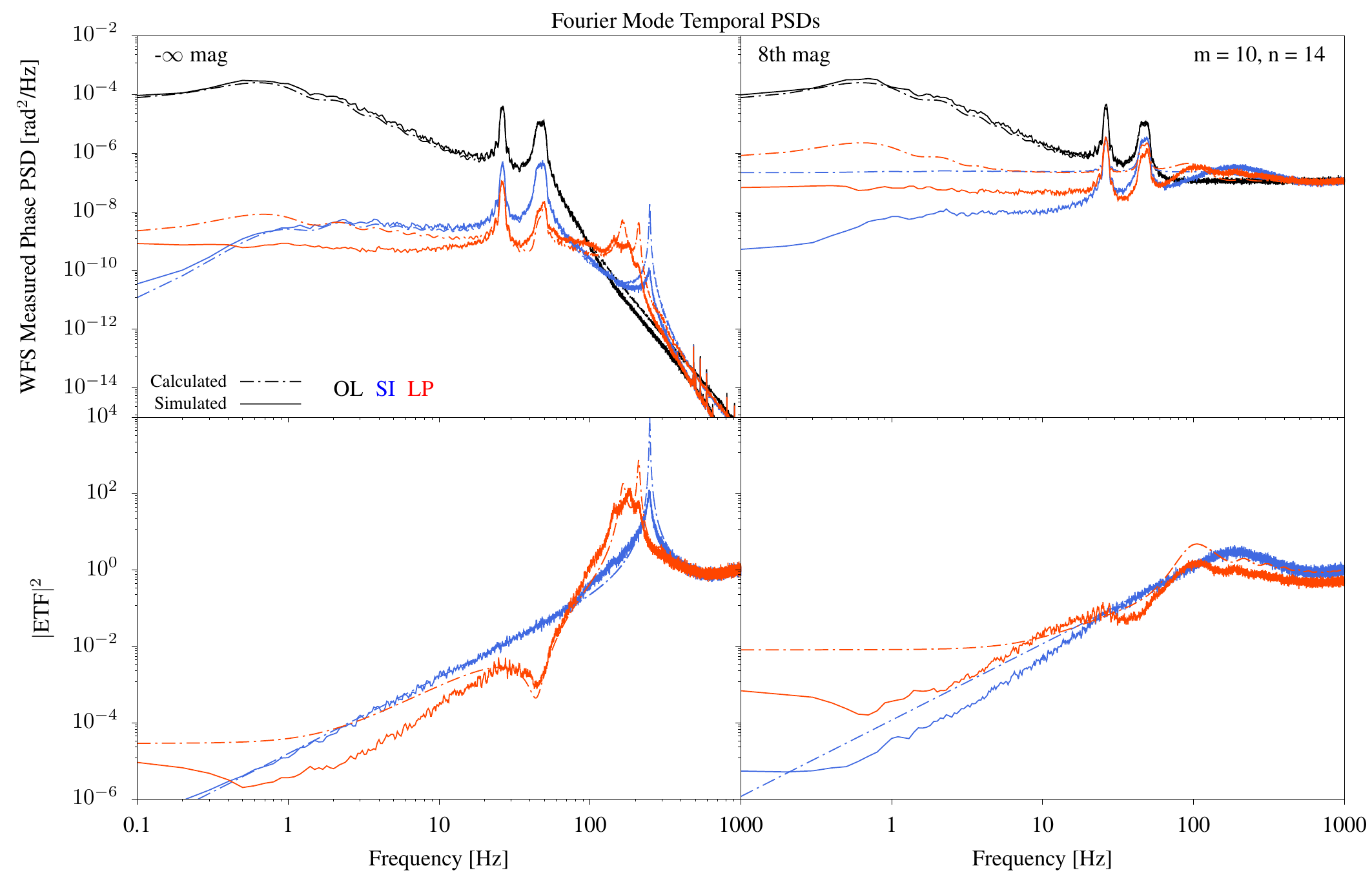}
\caption{Comparison of predicted and simulated PSDs for Fourier mode (m=10,n=14).  Due to the many subtle differences between the idealized semi-analytic model and the discretized end-to-end simulation, we do not expect perfect agreement.  See text for discussion.    \label{fig:opd_psdcomp}}
\end{figure}

Figure \ref{fig:opd_psdcomp} evaluates the dynamics of a single spatial frequency component  $m,n = 10,14$.  Note that this required re-fitting the Fourier coefficient after the simulations completed, since the control basis was different.  For consistency and brevity we use only this mode for illustration of single-mode dynamics.  It was arbitrarily chosen, and is representative of typical results.

In the top row, the black lines compare the calculated (dot-dashed) and simulated (solid) OL PSD.  The blue lines compare the output of the calculations to the simulations for the SI, and red lines show the LP.  The bottom row shows the ETF for each controller.  On the $-\infty$ mag guide star, the OL PSD shows excellent agreement at low frequencies, with a slight divergence in slope beginning at a $\sim$100 Hz. The SI likewise shows nearly perfect agreement, with the exception of a lower overshoot peak in the simulation.  The LP calculations are qualitatively a good match to the simulations, showing the same general shape, but the simulation shows better rejection at lower temporal frequencies.  This better low-frequency performance is true on the 8th mag star for both SI and LP.  The difference in integrated power at low frequencies is small and does not appreciably impact final contrast, and as we show in the following section leads to good results in intensity.  

The disagreements evident in the figure are due to the differences between the idealized continuous-time semi-analytic calculations and the discretized simulations. In particular the empirical optimization of the control laws produces different dynamical behavior in the simulations compared to the model predictions Additionally, the use of an orthogonalized basis in the simulation produces different noise rejection characteristics on the 8th mag star.  

Overall the comparison of outputs presented here give us confidence that the semi-analytic model gives reasonable results, and can be used to analyze the closed-loop contrast performance and wavefront dynamics in a coronagraphic AO system.

\subsection{Speckle Lifetimes}

\begin{figure}
\hspace{-0.3in}
\centering
\includegraphics[width=6.5in]{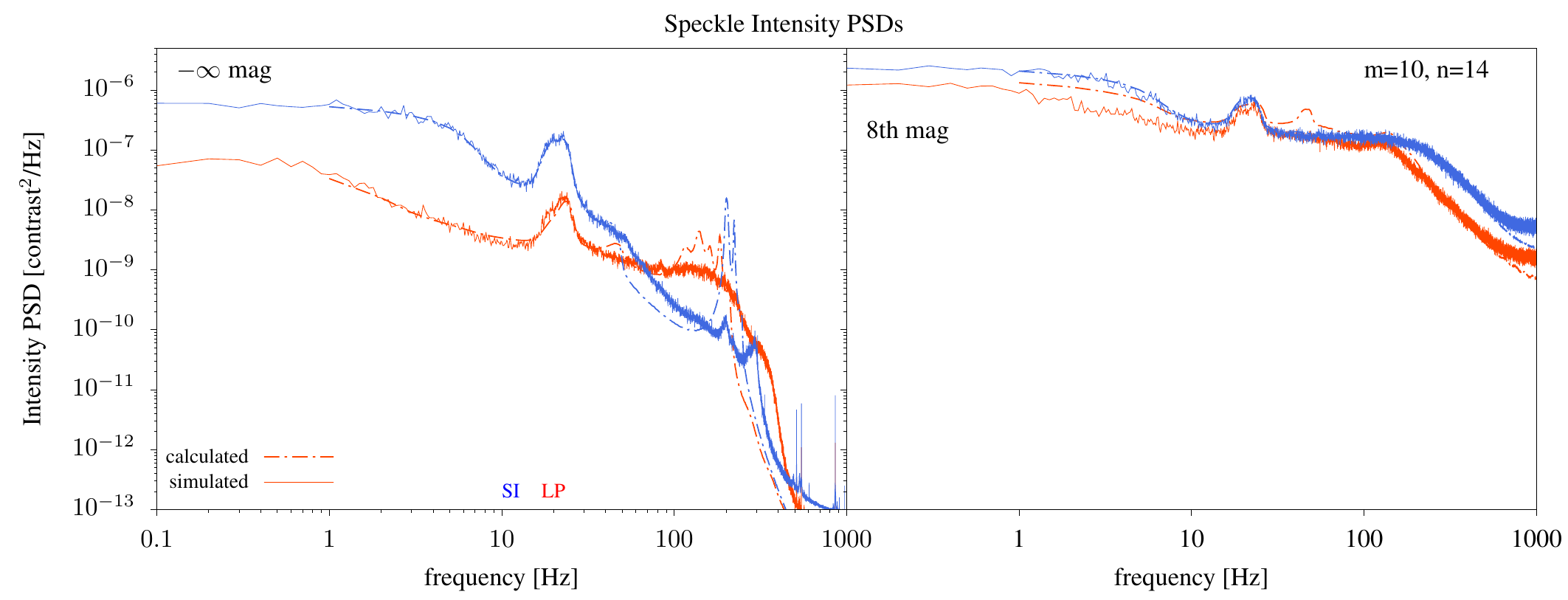}
\caption{Comparison of predicted and simulated PSDs for speckle intensity at position (m=10,n=14).  The agreement between the calculations and the simulations is very good at low frequencies, which determine the speckle lifetime.  At higher frequencies the calculations predict stronger peaks than occur in the simulations.  \label{fig:sppsdcomp}}
\end{figure}

In Figure \ref{fig:sppsdcomp} we compare the intensity PSDs in the post-coronagraph focal plane.  In intensity, the agreement between calculations and simulations is even better than in phase.  There is a departure at high frequencies, where the calculations predict stronger peaks than occur in the simulations. The agreements in overall power and the fraction of power contained in low-frequencies, which are most important for determining speckle lifetimes, are very good.

\begin{figure}
\hspace{-0.3in}
\centering
\includegraphics[width=6.5in]{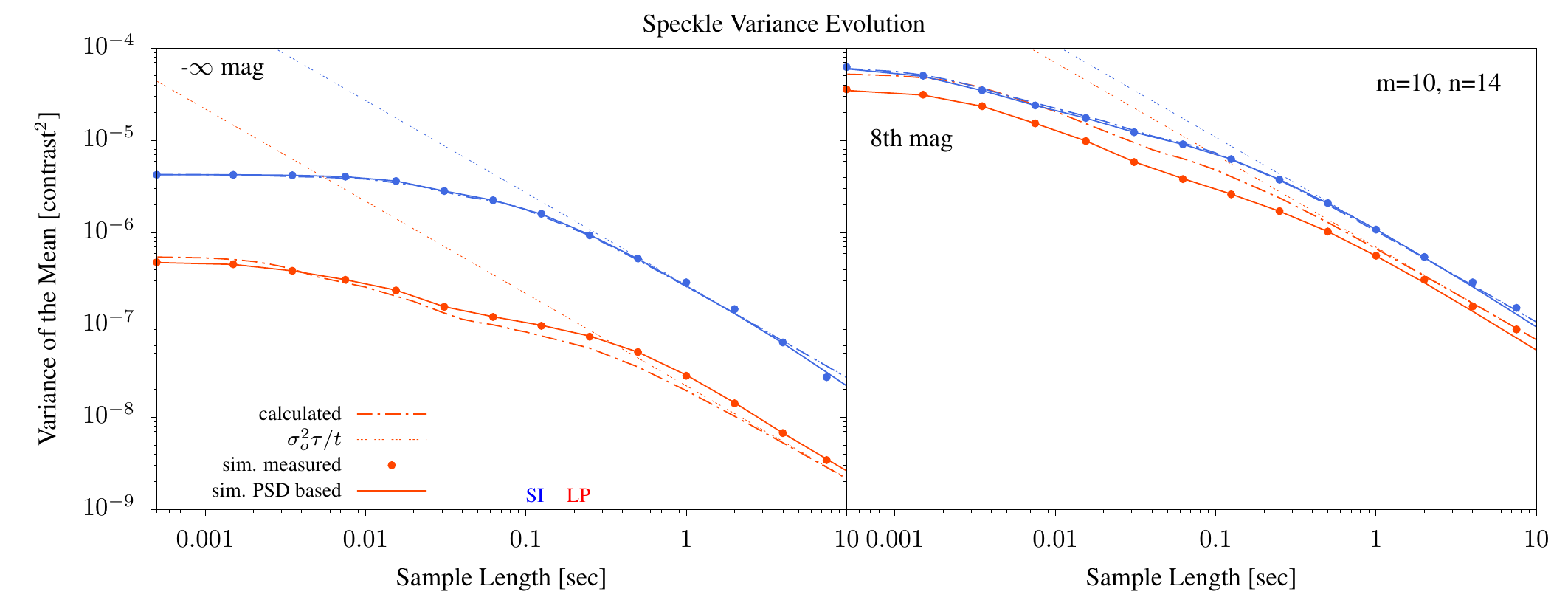}
\caption{Comparison of predicted and simulated variance evolution for speckle intensity at position (m=10,n=14).  Here ``calculated'' is the output of the method in Section \ref{sec:recipe}.  The ``sim. measured'' points (filled circles) show the direct measurements of variance as a function of time in the simulations.  The ``sim. PSD based'' lines show the results of using Equation (\ref{eqn:psd_lifetime}) to predict the variance as a function of time using the simulated intensity PSDs.  The comparison between the two simulation-based results shows that the analytic expression for speckle lifetime we derived is valid.  These are compared to the calculated line to show the result of the semi-analytic method for calculating the intensity PSD without an end-to-end simulation.  \label{fig:binvarcomp}}
\end{figure}

In Figure \ref{fig:binvarcomp} we show the evolution of the variance of the mean with time.  The lines show the result of applying Equation (\ref{eqn:varmean}) to PSDs of the intensity in the simulations. To test the validity of the derivation of speckle lifetime, the points show the direct calculation of the variance in the simulations, as a function of increasing averaging length.   Importantly, the solid lines are in nearly perfect agreement with the points, showing the validity of Equation (\ref{eqn:varmean}).  The dot-dashed lines, which are the output of the semi-analytic calculations, are slightly offset for the LP, which is due to differences in simulated vs. predicted contrast (equivalent to the overall variance level) as found in the previous section, but the shape of the predicted evolution with time for both SI and LP controllers is in good agreement with the simulations.

\begin{figure}
\hspace{-0.3in}
\centering
\includegraphics[width=3.54in]{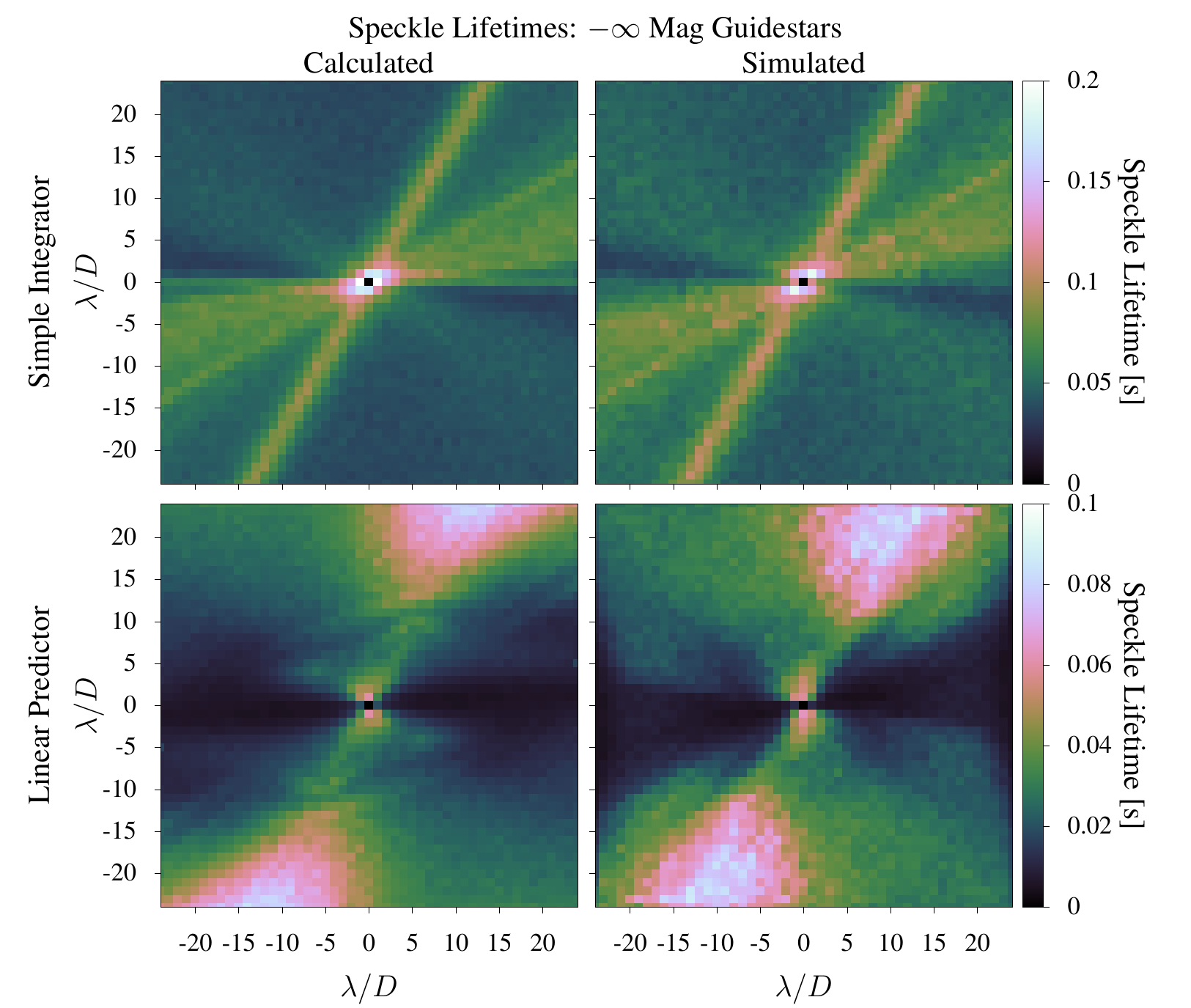}
\includegraphics[width=3.54in]{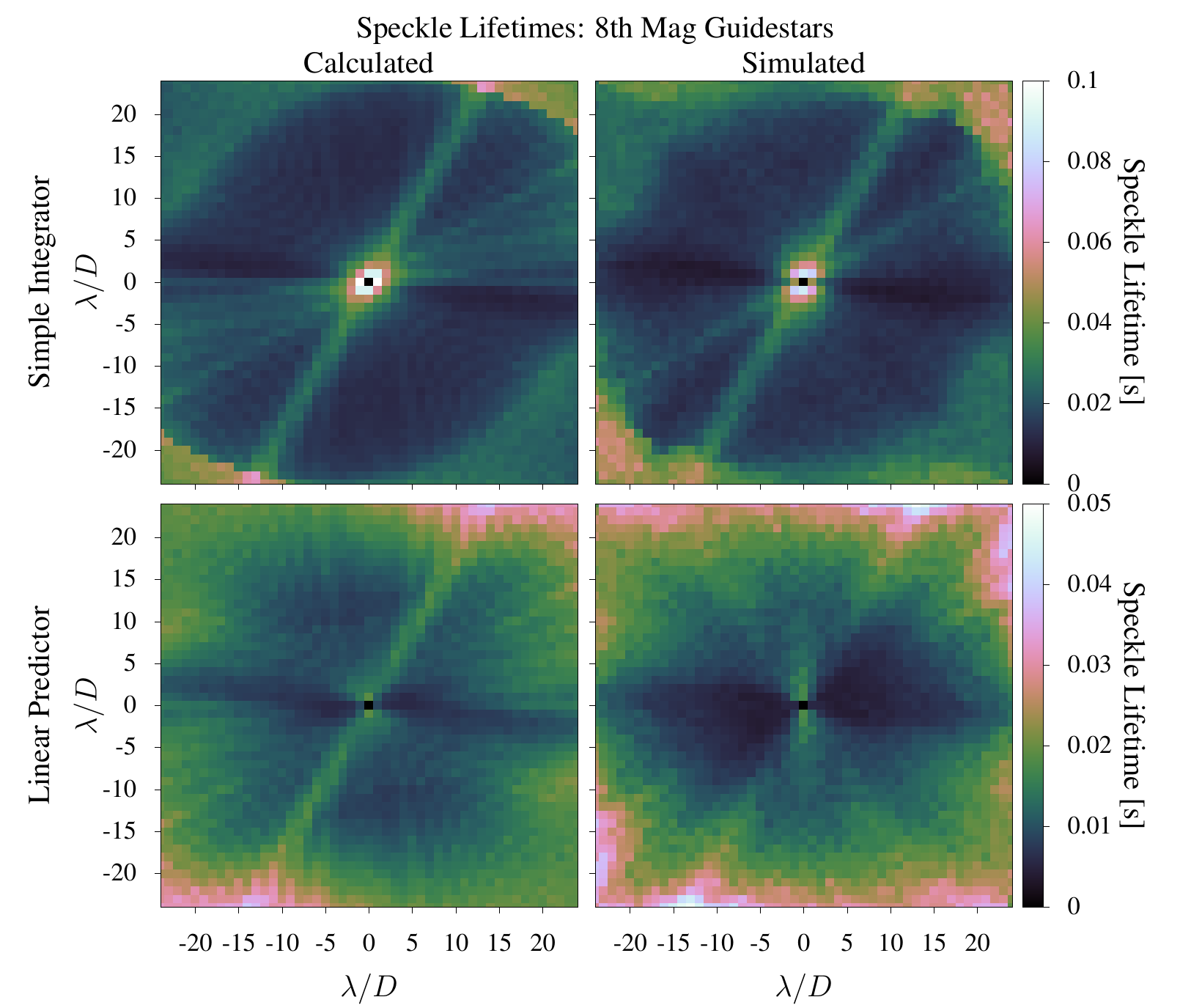}
\caption{Comparison of model predictions for speckle lifetime to end-to-end simulation for a 6.5 m telescope.  Left: without photon noise, considering the input turbulence only.  Right: with photon noise on an 8th mag star assuming the sensitivity of an unmodulated pyramid WFS. The same colorbar is used for both calculated and simulated within a combination of controller and guidestar, but is adjusted to best show detail for each combination. For the integrator the agreement is very good.  For the predictor on the infinitely bright star, the agreement is good overall but the calculations modestly over-predict speckle lifetimes in a wedge associated with (but not parallel to) the wind vectors.  These results demonstrate that AO control does affect speckle lifetimes, and that the lifetime is significantly shorter under predictive control.  The lifetime also depends on location in the control region. \label{fig:lifetime_2Dcomp}}
\end{figure}

\begin{figure}
\hspace{-0.3in}
\centering
\includegraphics[width=6.5in]{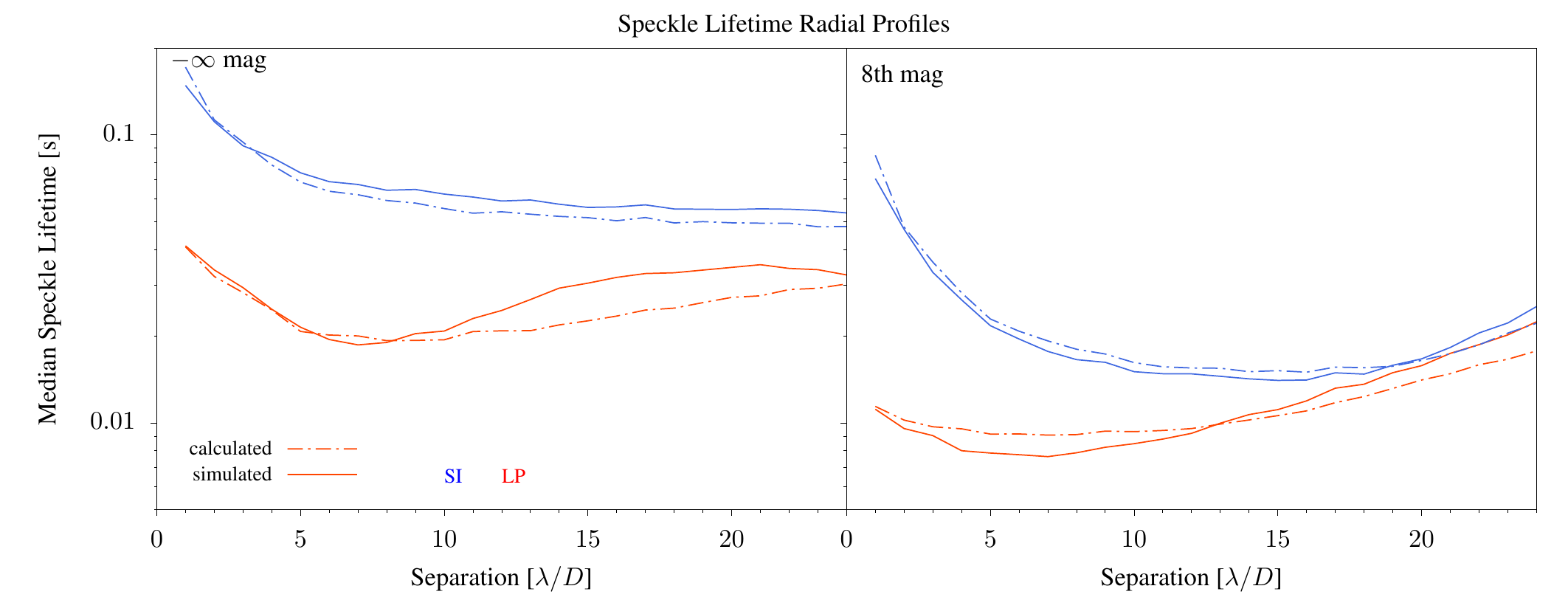}
\caption{Median radial profiles of the lifetime maps shown in Figure \ref{fig:lifetime_2Dcomp}.  Left: without photon noise, considering the input turbulence only.  Right: with photon noise on an 8th mag star assuming the sensitivity of an unmodulated pyramid WFS. Speckle lifetime is shorter under predictive control compared to the simple integrator. The large difference in median lifetime between simulation and calculation for the LP on the bright star is due to the wedge shaped region where the simulation predicts higher lifetimes (see Figure \ref{fig:lifetime_2Dcomp}, and text for discussion).  \label{fig:lifetime_radprof}}
\end{figure}

Finally in Figure \ref{fig:lifetime_2Dcomp} we present maps of speckle lifetime, and in Figure \ref{fig:lifetime_radprof} the median radial profile of speckle lifetime. The calculated and simulated lifetimes agree throughout most of the control region.  The noteworthy exception occurs in the LP roughly along a diagonal.  At wider separations the simulations predict lifetimes $\sim 50\%$ higher in a wedged shape region.  As with the differences in the phase dynamics discussed above, we attribute this to differences between the orthogonalized modal basis used in the simulations and the Fourier modes used in the calculations, and the resultant changes in the output of the controller coefficients and gain optimization.

For comparison, the ``crossing-time'' speckle lifetime of \citet{2005SPIE.5903..170M} would be $\tau=0.3D/\mathcal{V} = 224$ ms for the turbulence averaged wind speed used in these calculations, with the caveat that the constant of proportionality is expected to vary with coronagraph. We find lifetimes much shorter than this, up to a factor of 10 shorter for the LP controller. Figures \ref{fig:lifetime_2Dcomp} and \ref{fig:lifetime_radprof} show that speckle lifetime has a strong dependence on control law, and varies with position in the image.  We investigate the $D/\mathcal{V}$ scaling below.

Overall these results show that Equation (\ref{eqn:psd_lifetime}) correctly describes the statistical speckle lifetime for a given intensity PSD, which in turn can be predicted by our semi-analytic model of an ExAO coronagraph.

\clearpage

\section{Speckle Lifetime Comparisons}
\label{sec:trends}

We now investigate the behavior of speckle lifetime over a range of parameters.  The first comparison we make is between control laws.  Here we simply calculate a histogram of the lifetimes over the 48$\times$48 $\lambda/D$ dark hole for the SI and LP controller.  We also apply the Monte Carlo procedure to the OL case, by treating the input PSDs as unmodified by control laws.  We note that this is not strictly valid as the input phase is then outside the approximation regime assumed in the derivation of Equation (\ref{eqn:cross_terms}).  However it provides a useful comparison of the change in the distribution of power in the PSD caused by closed-loop control action. 

\begin{figure}[h]
\centering
\includegraphics[width=3.25in]{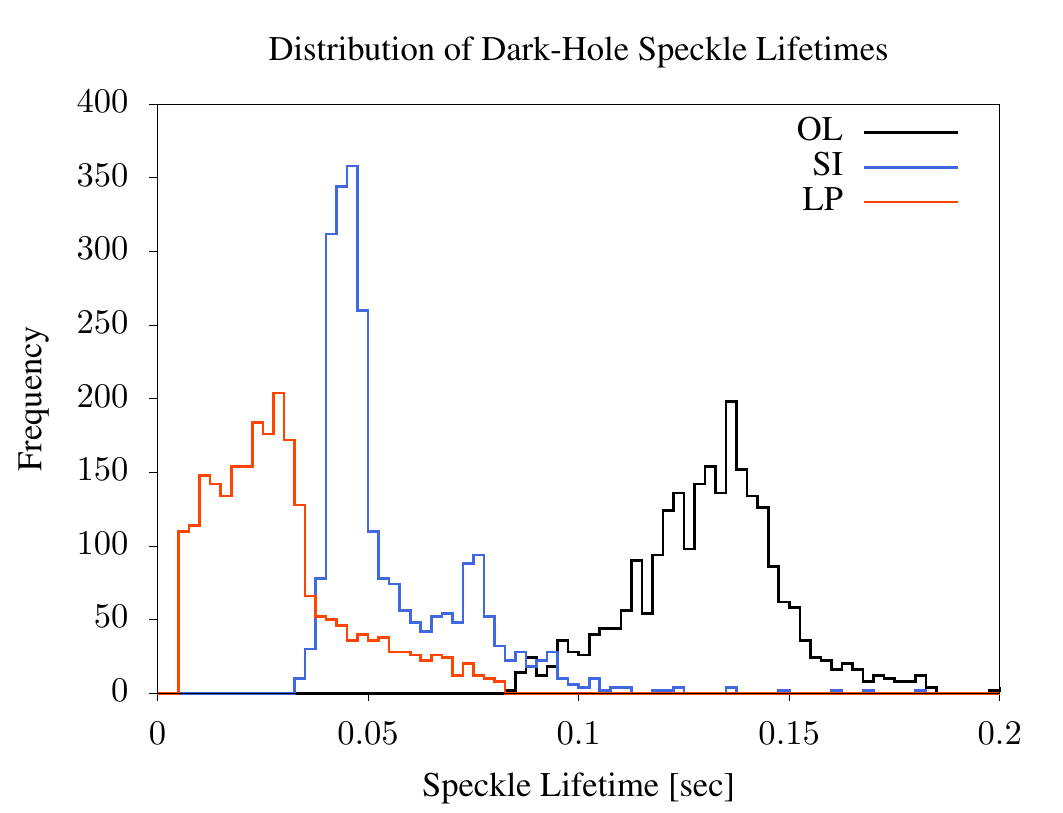}
\caption{Distribution of speckle lifetimes in the dark-hole for open-loop (OL), simple integrator (SI), and linear predictor (LP) controllers. Each histogram contains 2400 measurements.  AO control shows a strong impact on speckle lifetime, with predictive control improving further over that of the simple integrator. 
\label{fig:slHist}
}
\end{figure}

The three resulting histograms are shown in Figure \ref{fig:slHist}.  It is clear that AO control shortens the speckle lifetime, and that predictive control further reduces speckle lifetime compared to the integrator.

\begin{figure}
    \includegraphics[width=3.5in]{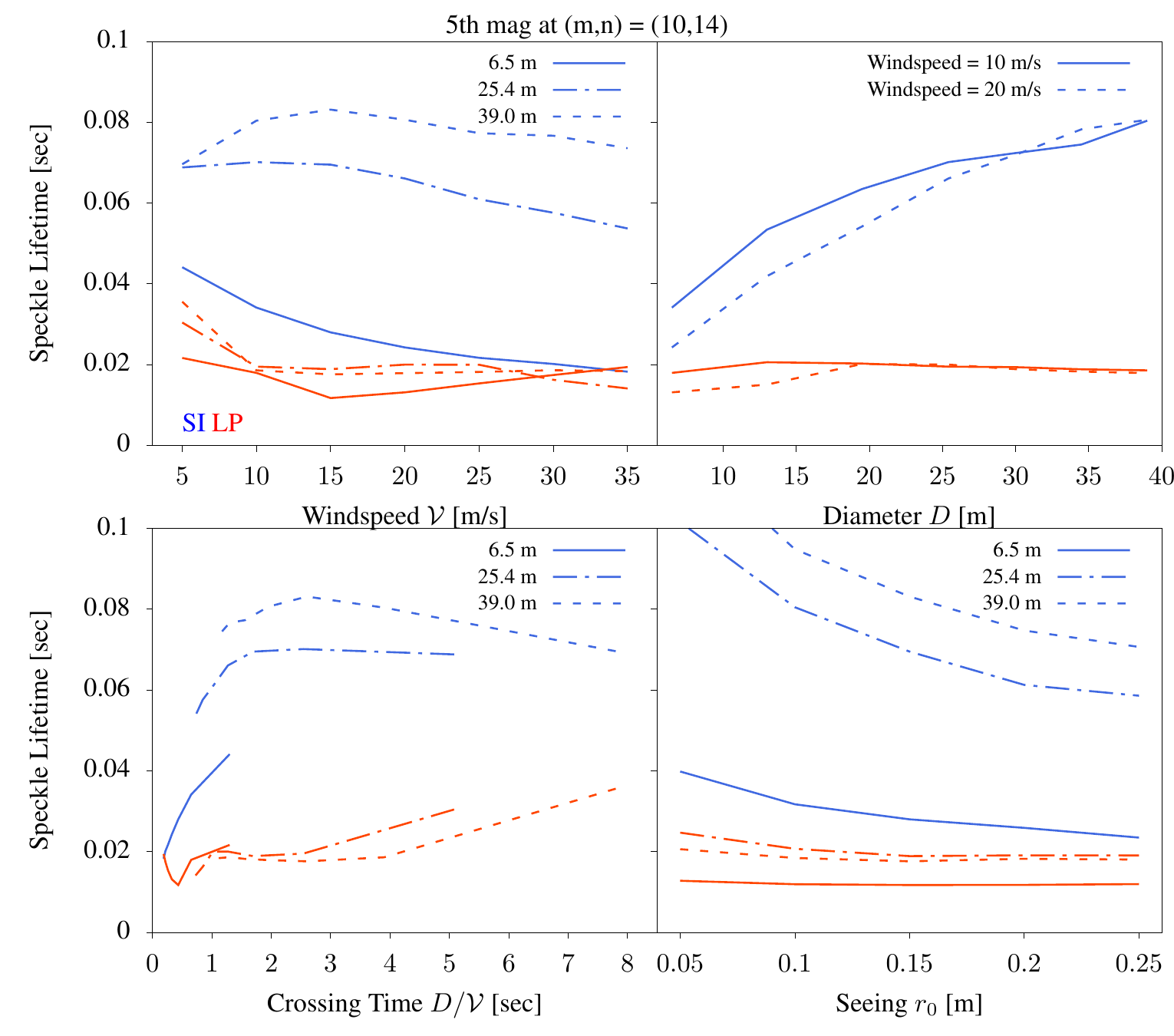}
    \includegraphics[width=3.5in]{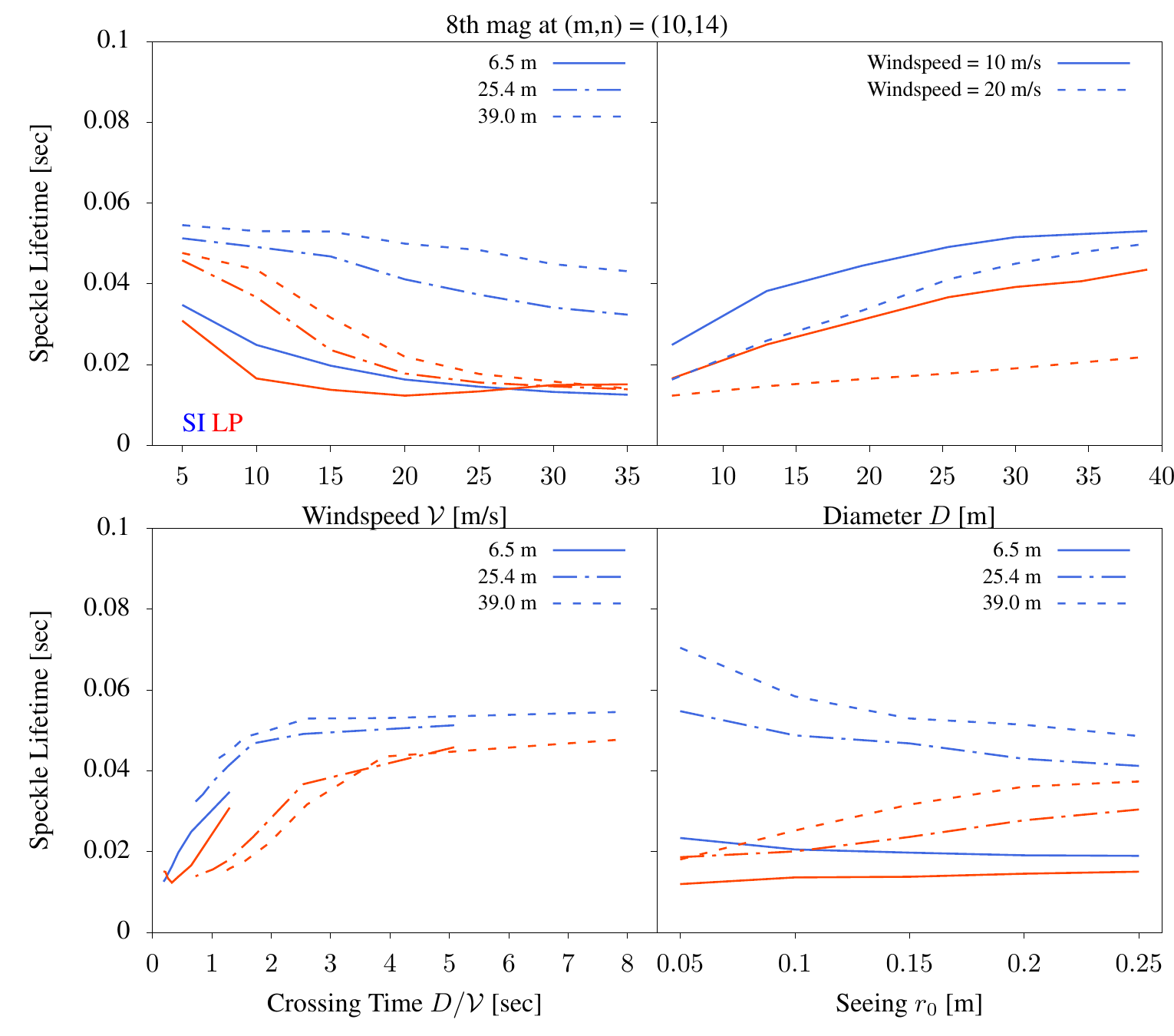}
    \caption{Comparison of speckle lifetimes over a range of parameters.  We compare speckle lifetimes with a 5th mag guide star (left) and 8th mag guide star (right) on telescopes of diameter $D$ up to 39 m, with wind speeds $\mathcal{V}$ up to 35 m/s, and seeing $r_0$ up to 0.25 m.  Only the speckle lifetime for the $(m,n)$ = $(10,14)$ mode is analyzed here, for a 2 kHz AO system.  For the simple integrator (SI, blue), speckle lifetime generally decreases on fainter stars due to the larger contribution of photon noise compared to the turbulence power in the control system.  Additionally the SI shows some dependence on the crossing time up to $D/\mathcal{V}\sim2$.  The linear predictor (LP, red) controller is somewhat less dependent on these parameters.  These comparisons demonstrate that AO control has a large impact on speckle lifetime, and that there are no simple thumb rules which can be used to predict speckle lifetime.  \label{fig:scalings}}    
\end{figure}

Next, we analyze the change in speckle lifetime with guide star brightness, telescope diameter, wind speed, and seeing.   We held relative spatial frequency constant with $D$, such that the resultant speckle occurs at the same relative projected separation in $\lambda/D$ units.  This compares the performance of coronagraphic AO systems given the improvement in spatial resolution with $D$.  Note that the input power per Fourier mode scales as $D^{-2}$, that is it will be lower given larger $D$.  This means that the input temporal PSD has a different shape due to relative spatial frequency as well as different power due to both the wind speed and diameter. The input noise PSD will also be lower for each mode.  The balance between these inputs can be expected to affect the performance of the optimized controllers, resulting in variation in speckle lifetimes.

Figure \ref{fig:scalings} compares speckle lifetimes for a 5th-magnitude guide star to those for an 8th-mag guide star.  The lifetime is generally shorter on the 8th-mag star due to the increased influence of photon noise relative to the turbulence PSD, with the overall trends being similar.  Lifetimes are shorter under LP control compared to SI control for both guide stars for all parameters.

The SI controller, in general, shows a decrease in lifetime with wind speed $\mathcal{V}$ and an increase with diameter $D$ for both guide stars.  When analyzed as the crossing time $D/\mathcal{V}$, the SI controller scales roughly proportional to the crossing-time up to $D/\mathcal{V}$ $\sim$ $2$.  For longer crossing times the trend becomes flat.  This implies that the control system becomes dominated by photon noise rather than servo lag and so the optimization process no longer produces changes in the output dynamics.

The LP controller shows much weaker dependence on the same parameters.  Except at the lowest wind speeds, the speckle lifetime is $\sim$$20$ ms on the 5th-magnitude guide star with little variation.  Here the photon noise is well below the input turbulence PSD, and the LP can effectively whiten the disturbance to the photon noise floor.  The result is that the output PSD has the same shape, giving the same speckle lifetime regardless of the parameters.  On the 8th-magnitude star photon noise makes predictive control less effective, causing some variation with the parameters. There is a proportional scaling up to $D/\mathcal{V}$$\sim$$3$. 

For the SI controller, improving seeing (larger $r_0$) reduces speckle lifetime.  As the turbulence power is reduced, the control system is better able to reject low temporal frequencies, resulting in lower speckle lifetimes.  As with $D$ and $\mathcal{V}$, the LP controller shows little variation in lifetimes vs. $r_0$.  Here again, the predictive controller whitens the noise to the same floor set by photon noise, which produces an output PSD of roughly the same shape resulting in similar speckle lifetimes.  The lifetime increases somewhat vs $r_0$ on the larger diameters as the turbulence power becomes relatively lower compared to photon noise, resulting in reduced effectiveness of the predictor such that more correlations remain in the output.

\section{The Potential of PSF Reconstruction Post-Processing}
\label{sec:reconstruction}
The reconstruction of the post-AO PSF using AO system telemetry has been studied for some time, primarily focused on the long-exposure PSF, e.g. \citet{1997JOSAA..14.3057V} and see the recent review by \citet{2020SPIE11448E..0AB}.  PSF reconstruction in the context of high-contrast imaging is typically based on image data alone \citep{2006ApJ...641..556M,2007ApJ...660..770L}, however the potential for telemetry-based PSF estimation in high-contrast imaging has been recognized \citep{2013ApJ...767...21F,2013ApJ...767..100C}. Here we use our treatment of the intensity dynamics behind an AO-fed coronagraph to analyze the performance of PSF estimation and subtraction.  We are particularly motivated by the use of WFS telemetry to improve on the sensitivity limit set by atmospheric speckle lifetimes, however this analysis could be applied to conventional post-processing such as the KLIP algorithm \citep{2012ApJ...755L..28S}.

A closed-loop wavefront control system measures the coefficients of the controlled modes in the residual wavefront at each loop time step.  The WFS produces measurements of the residual phase with modal coefficients $\Delta h_{mn}^{p}(t)$ which we assume are recorded. Additionally we assume that focal plane intensity is recorded with a detector reading out at the same rate as and synchronized to the WFS, such that measurements $I_{mn}(t)$ are recorded at each pixel. The measured focal plane intensity may include signal from a planet in addition to scattered starlight.

At some later time we use the measured residual coefficients to reconstruct an estimate of the focal plane intensity due to scattered starlight alone.  To begin, we reconstruct the residual phase based on Equation \ref{eqn:phi_expansion} 
\begin{equation}
\Delta\Phi(\vec{q},t) = \frac{2\pi}{\lambda} \sum\limits_{mn} \left[ \Delta h_{mn}^{+}(t) M_{mn}^{+}(\vec{q}) + \Delta h_{mn}^{-}(t) M_{mn}^{-}(\vec{q})\right].
\label{eqn:phi_residual}
\end{equation}
We then propagate the residual phase to the focal plane
\begin{equation}
\hat{I}(\vec{r},t) = G \left\{\mathcal{A}(\vec{q}) \mbox{e}^{i\Delta\Phi(\vec{q},t)} \right\}
\end{equation}
where $\mathcal{A}(\vec{q})$ defines the pupil and $G\{\cdot\}$ is the system propagation operator, which could include information such as the effect of the coronagraph, non-common path aberrations (not sensed by the WFS which measures the $\Delta h_{mn}^{p}(t)$), and Fresnel propagation.  See \citet{2017MNRAS.467L.105H} for an example of the construction of such a model. In the case of the perfect coronagraph and ideal optical system used in this work, the propagation operator reduces to Equation \ref{eqn:cross_terms}. The result of applying $G\{\cdot\}$ to the reconstructed wavefront is an estimated intensity time series at position $\vec{r}_{mn}$ 
\begin{equation}
\hat{I}_{mn}(t) = \hat{I}(\vec{r}_{mn},t)
\end{equation}

\begin{figure}
\centering
\includegraphics[width=4in]{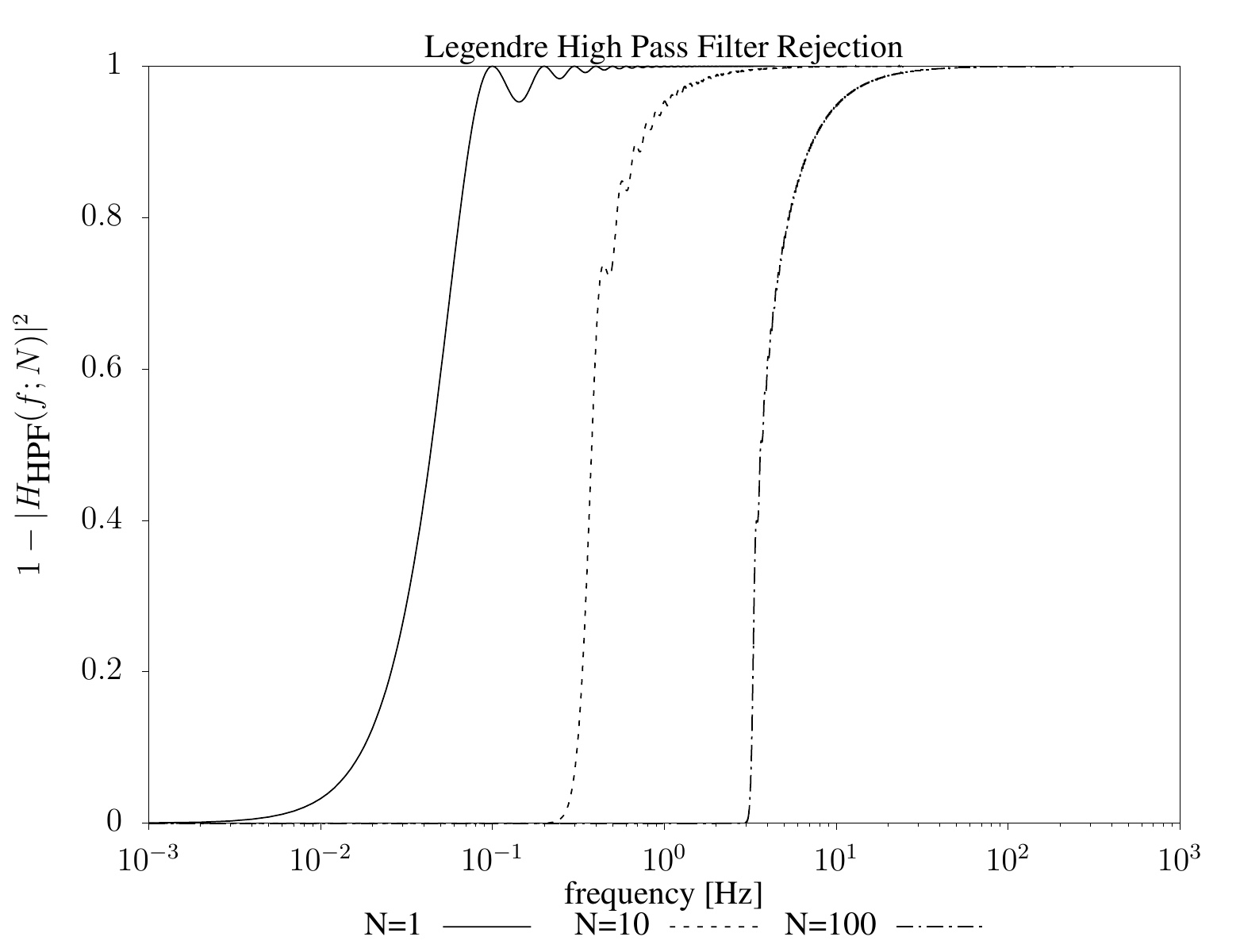}
\caption{The Legendre high pass filter rejection functions for 1, 10, and 100 modes. Fitting and subtracting Legendre polynomials reduces low-frequency power, which lowers speckle lifetimes.\label{fig:leghpf}}
\end{figure}

In Section \ref{sec:psd_lifetime} we developed the expansion of the intensity time-series in terms of orthonormal Legendre polynomials $P_\eta$.  For a segment of the reconstructed intensity of length T, we measure the coefficients of the Legendre polynomials
\begin{equation}
\hat{a}_\eta = \int_{-1}^1 \hat{I}_{mn}\left(\frac{T}{2}\xi\right) P_\eta(\xi) d\xi. 
\end{equation}
We then subtract the first $N$ Legendre modes of the reconstructed time-series from the measured focal plane intensity time-series, giving the ``post-processed'' intensity
\begin{equation}
I_{mn,pp}(t) = I_{mn}(t) - \sum_{\eta=0}^{\eta=N-1} \hat{a}_\eta P_\eta\left(\frac{2}{T}t\right). 
\end{equation}
The outcome of this post-processing will be fewer long-term correlations as the low order modes are successively subtracted, which will in turn lower the speckle lifetime.  We can describe this outcome statistically as a high-pass filter (HPF).  The PSD of a mode is
\begin{equation}
\mathcal{T}_\eta(f) = \mathcal{T}(f) \left|Q_\eta(f)\right|^2
\end{equation}
where ${Q}_\eta$ is the Fourier transform of $P_\eta$, which we derive in Appendix \ref{app:tau_psd}, giving
\begin{equation}
 \left|Q_\eta(f)\right|^2 = \left(\frac{2\eta+1}{2}\right)\frac{J_{\eta+\frac{1}{2}}^2(\pi T f)}{T f}.
\end{equation}
After fitting and subtracting $N$ modes, the speckle intensity PSD will be high-pass filtered with squared transfer function
\begin{equation}
\left| H_\mathrm{HPF}(f;N) \right| ^2 = 1 - \frac{1}{2}\sum_{\eta=0}^{\eta=N-1} (2\eta+1) \frac{J_{\eta+\frac{1}{2}}^2(\pi T f)}{T f}.
\end{equation}
We show examples of the transfer functions for this filter in Figure \ref{fig:leghpf}.  

Now due to measurement noise the fit to the data will have a fractional uncertainty $\gamma$, so the resulting speckle intensity PSD will be
\begin{equation}
\mathcal{T}_{I_{mn},pp}(f;N,\gamma) = \left( 1 -  \frac{(1-\gamma)^2}{2}\sum_{\eta=0}^{\eta=N-1}  (2\eta+1) \frac{J_{\eta+\frac{1}{2}}^2(\pi T f)}{T f}\right) \mathcal{T}_{I_{mn}}(f).
\end{equation}

\begin{figure}
\centering
\includegraphics[width=4in]{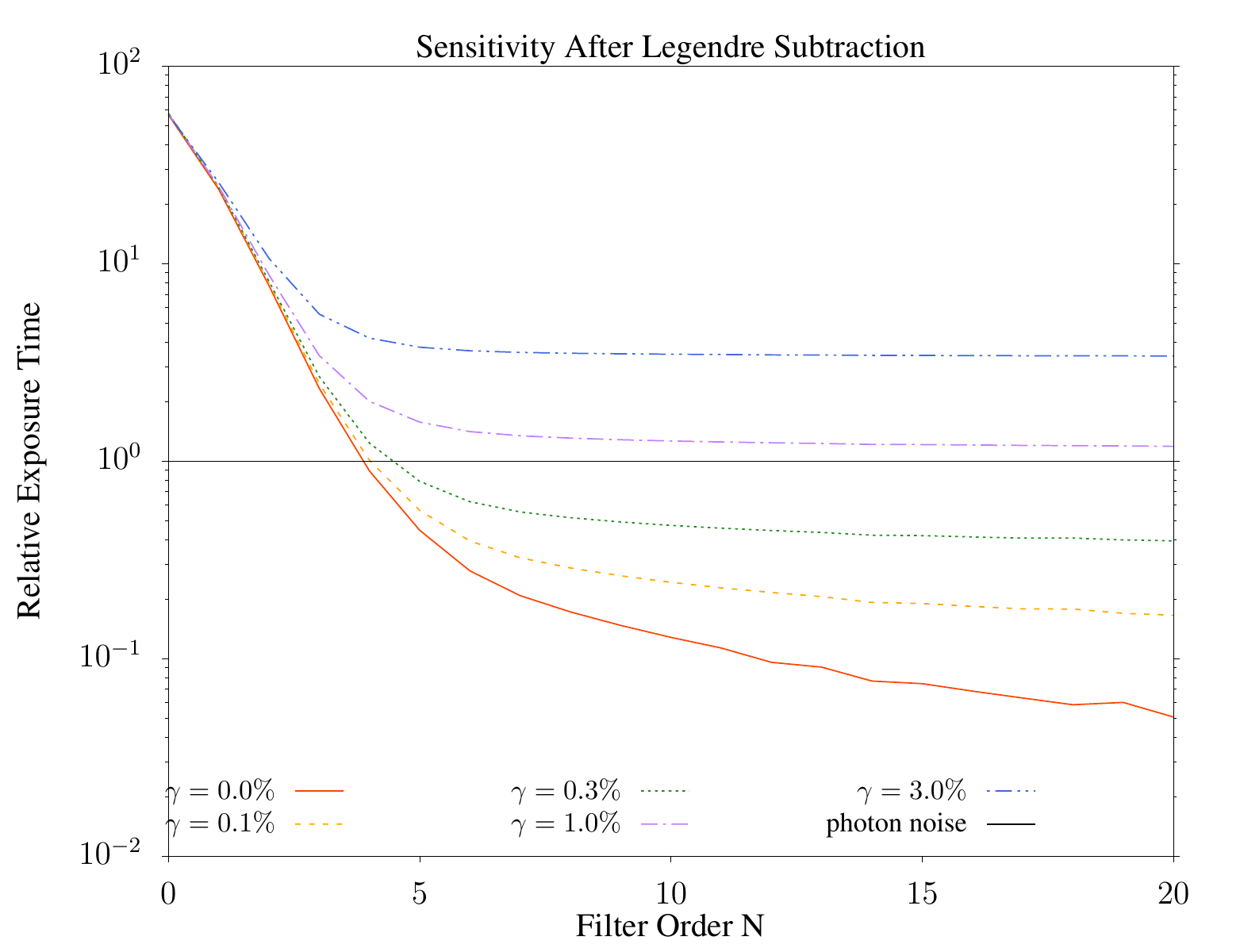}
\caption{The relative sensitivity, in terms of exposure time, after subtracting the first $N$ Legendre polynomials with uncertainty in the fit $\gamma$ for each $a_\eta$.  The horizontal line at 1 corresponds to the photon-noise limited exposure time on an 8th mag star.  Residual-atmospheric speckles increase the exposure time by a factor 55 compared to the photon-noise limit.   
WFS telemetry-based PSF subtraction of just the first $\sim$5 low-order Legendre polynomials from the speckle time series reduces the speckle lifetime, improving sensitivity dramatically. If better than $\sim$$1\%$ accuracy in reconstruction of these modes can be achieved, the speckle noise can be reduced to below the photon noise limit. \label{fig:legrelsens}}
\end{figure}

In Figure \ref{fig:legrelsens} we show the improvement in exposure time as defined by Equation (\ref{eqn:relexptime}), after fitting and removing the first $N$ Legendre polynomials, with an error of $\gamma$ on the fit.  In the figure, the horizontal line corresponds to the photon-noise limited exposure time on an 8th magnitude star.  If the first $\sim$5 low-order Legendre polynomials can be determined from, say, WFS-based reconstruction, the reduction in speckle lifetime from removing the low-order temporal modes improves sensitivity significantly.  If the modes are reconstructed to better than $\sim$1\% accuracy in the $a_\eta$, the photon-noise limit in the focal plane can be reached.

\section{Discussion}
\label{sec:discussion}

\subsection{Caveats}

The results shown here are subject to several caveats.  All of the assumptions used in the development of \citet{2018JATIS...4a9001M} apply here.  In particular, our analysis is monochromatic. While we assumed that the science and sensing wavelength were the same, neither component of an AO system can be truly monochromatic.  Though speckle lifetime formally does not depend on wavelength, which only scales the PSD without changing its shape, AO performance will depend on how light is distributed in the instrument.  Furthermore, atmospheric chromaticity will cause differential errors across the bandpass of the WFS, and between the science and WFS bandpasses.  These errors will contribute speckles in addition to those explicitly treated here, which will have different dynamics as they are not subject to the same control filters.  These speckles can, however, be analyzed with the same tools we developed here.  We save this for future work.

A key assumption underlying the mathematics of the PSD analysis is that the speckle intensity is wide-sense stationary (WSS), which means that the mean is constant and that the autocorrelation does not depend on time.  As discussed above, this assumption implies that $r_0$ and the layer $\vec{\mathcal{V}}$ are constant such that the statistics of the AO control output process are WSS.  In the simulations presented these assumptions are true by construction, and so the comparison to our semi-analytic predictions is valid.  On-sky input turbulence parameters will vary, and so the process will not be WSS over long periods.  However, these variations occur over finite times.  For our results to be useful the time-scales of these variations would need to be long enough to estimate a PSD which meaningfully captures the intensity dynamics.   What this ultimately means is that we must consider the speckle lifetime as itself a time-variable parameter, but we can expect there to be a representative value over an observation that characterizes long exposure sensitivity.

\subsection{The Semi-analytic Model}

We have developed a semi-analytic model of a coronagraphic AO system that allows us to estimate long-exposure contrast, the intensity dynamics, and statistical lifetimes.  Compared to an idealized end-to-end simulation, the results of the model compare favorably.  This powerful technique allows us to analyze the impact of control-law choice, as well as to compare the influence of various telescope and atmosphere parameters on the speckle lifetime.  Key benefits of such modeling are that the true long-exposure performance of a system can be determined without the expense of simulations.  For instance, the parameter study in Section \ref{sec:trends} would have been prohibitively expensive if conducted with simulations. As design efforts continue for the coming GSMTs, such models will be useful for similar parameter studies to optimize high contrast imaging instruments.

\subsection{The Impact of AO}

We analyzed various parameters with potential impacts on speckle lifetime.  Overall, the simplest summary of this analysis is that there is no single thumb-rule-type relationship that applies to all parameters.  This modestly contradicts the conclusions of \citet{2005SPIE.5903..170M}.  There, the AO system was modeled as a spatial filter, with the result that while the speckle amplitude was reduced, the evolution of correlations depended only on the crossing-time.  As a result, neither the control action nor the overall power (seeing) should impact speckle lifetime.  Only the wind would change speckle lifetime for a given pupil diameter.

Here we have added the time dimension, fully considering the action of AO control as a temporal filter.  Furthermore, we have modeled the impact of high-order predictive control.  With this more complete view, we find that the control action does reduce speckle lifetime.  There is an influence of the crossing-time as predicted by \citet{2005SPIE.5903..170M} to some extent, but in addition to the reduction in amplitude, the time-domain influence of the control system is to increase the rate at which one speckle turns into the next.  That is, the control reduces the correlation length.  This is consistent with the findings of \citet{2006OExpr..14.7499P}, which predicted that control laws could be optimized based on speckle lifetime to maximize sensitivity.

Predictive control, already expected to improve contrast \citep{2007JOSAA..24.2645P,2017JOSAA..34.1877C,2018JATIS...4a9001M, 2021arXiv210307566H}, will further improve sensitivity by decreasing the speckle averaging lifetime.  On bright stars, predictive control nearly eliminates dependence on the crossing-time as it acts to reduce servo-lag errors.

As ExAO coronagraphs approach the residual atmosphere limit, it will be increasingly important to understand the impact of correlated noise from atmospheric speckles on sensitivity.  Our results demonstrate that this is a complicated task, that must consider all relevant parameters of the system under study.

\subsection{Post-Processing}
The temporal decorrelation of speckle noise is one of the main benefits of PSF-subtraction post-processing \citep{2006ApJ...641..556M,2008ApJ...673..647M}, even though algorithms have typically focused on exploiting spatial correlations.  Recently, studies have begun to consider the time-domain in post-processing explicitly \citep{2021A&A...646A..24S}, and several algorithms show promise in exploiting short exposure focal-plane data \citep{2019PASP..131k4506W, 2021arXiv210306898S} and WFS telemetry \citep[][submitted]{2021arXiv210506589R}\citep[][submitted]{2021arXiv210506590F} to estimate speckle intensity.   Knowledge of the intensity PSD allows for such algorithms to be analyzed efficiently.  We demonstrate this by considering the prospect of telemetry-based reconstruction of the speckle time-series. The algorithm we outlined made use of the main WFS images, but other information sources such as accelerometers, low-order WFS telemetry, and environmental sensors could be used.  The key challenge is in the construction of the system propagation model.  We used the temporal intensity PSDs to predict the sensitivity gain possible if this can be done.  If the first few low-order temporal modes, here described in terms of the Legendre polynomials, can be determined and subtracted from the intensity in the focal plane, large improvements in exposure time are possible.  The fact that only a few low-order temporal modes need to be reconstructed has the important consequence that any such post-processing algorithm does not need to accurately determine the intensity at every time interval, and does not need to be perfect.  Accuracy in the mean, slope, and other low-order terms of a few percent is all that is needed to lower the speckle lifetime and therefore increase sensitivity.

\section{Conclusion}
The residual atmosphere after AO correction sets the fundamental limit for ground-based high-contrast imaging.  They key parameter for understanding this limit is the statistical speckle lifetime.  We have developed a semi-analytic method for calculating the speckle lifetime, and verified the model outputs using end-to-end simulations.  We have shown that the speckle lifetime is an order of magnitude shorter than previous studies have found, and analyzed the behavior of speckle lifetime with various telescope parameters.  The key development of the post-coronagraph intensity PSD allowed us to analyze post-processing as a time-domain filter and show that it is possible to achieve photon-noise limited sensitivity given a suitable reconstruction algorithm. In future work we will use this framework to analyze more realistic turbulence models, extend these techniques to instrumental quasi-static aberrations, and analyze various post-processing strategies.  \\

\noindent Acknowledgements: JRM and OG acknowledge support from the NSF (\#1625441, MagAO-X).  JRM, MPF, and OG acknowledge support from the NSF (\#1710356) and from the Heising Simons Foundation.

\appendix
\section{The Noise Lifetime of a 1-D PSD}
\label{app:tau_psd}

In one of the seminal analyses in the development of AO, \citet{1976JOSA...66..207N} showed how to calculate the covariance of the Zernike polynomials in Kolmogorov turbulence, using only the spatial power spectral density of the optical phase. Here we adapt that analysis to one-dimensional PSDs of arbitrary form.  

As above, we have a one-dimensional, one-sided ($f \ge 0$),  PSD $\mathcal{T}(f)$, where $f$ is normalized on $0 \leq f \leq f_s/2 = 1/(2\Delta t)$ according to Equation (\ref{eqn:process_var}).  We obtain a time-series $I(t)$ from the process governed by $\mathcal{T}(f)$, which we expand in terms of normalized Legendre polynomials defined in Equation (\ref{eqn:basis}).  The orthogonality condition is given in Equation (\ref{eqn:orthonormal}).
 
\subsection{Fourier Transform}

The Fourier transform of the Legendre polynomials is\footnote{See \url{https://dlmf.nist.gov/18.17\#v} 18.17.19}
\begin{equation}
\int_{-1}^{1} \mathscr{P}_\eta(x) e^{i2\pi k x} dx  =  i^\eta \frac{J_{\eta+\frac{1}{2}}(2\pi k)}{\sqrt{k}}
\end{equation}
with $J_\eta$ denoting the Bessel functions of the first kind.  We can then write the Fourier transforms of the orthonormal $P_\eta(x)$ as 
\begin{equation}
Q_\eta(k) = i^\eta \sqrt{\frac{2\eta+1}{2}}\, \frac{J_{\eta+\frac{1}{2}}(2\pi k)}{\sqrt{k}}.
\end{equation}
For analysis of Legendre mode filtering these have the useful property that\footnote{See \citet{2007tisp.book.....G} 6.574 \#2.}
\begin{equation}
\int_{0}^\infty \left| Q_\eta(k) \right|^2 dk = 1 \mbox{ for all }\eta.
\end{equation}

\subsection{Noll Analysis}

The expansion of $I(t)$ in terms of the $P_\eta(\xi)$ is given by Equations (\ref{eqn:basisexpansion}) and (\ref{eqn:expansecoeff}).  This series expansion in terms of normalized Legendre polynomials is the 1-D analog of the expansion in Zernike polynomials used by \citet{1976JOSA...66..207N}.  Following that analysis we write the covariance between any two $P_\eta(\xi)$ as
\begin{equation}
<a_\eta^* a_{\eta'}> = \int \int P_\eta(\xi) C\left(\frac{T}{2}\xi, \frac{T}{2}\xi' \right) P_{\eta'}(\xi) d\xi d\xi' 
\end{equation}
the autocorrelation is given by
\begin{equation}
C\left(\frac{T}{2}\xi, \frac{T}{2}\xi' \right) = \left<I\left(\frac{T}{2}\xi\right)I\left(\frac{T}{2}\xi'\right) \right>.
\end{equation}
Equivalently, in the Fourier domain we have
\begin{equation}
<a_\eta^* a_{\eta'}> = \int \int Q_\eta^*(\kappa) \Phi \left(\frac{2}{T}\kappa, \frac{2}{T}\kappa' \right) Q_{\eta'}(\kappa) d\kappa d\kappa' 
\end{equation}
where for a wide-sense stationary process
\begin{equation}
\Phi \left(\frac{2}{T}\kappa, \frac{2}{T}\kappa' \right) = \frac{2}{T} \mathcal{T}\left(\frac{2}{T}\kappa \right) \delta(\kappa-\kappa')
\end{equation}
after the change of variables $f \rightarrow (2/T)\kappa$ in the PSD $\mathcal{T}(f)$.

We now have the covariance of the Legendre coefficients
\begin{equation}
\langle a_\eta^* a_{\eta'}\rangle = i^{\eta'-\eta} \frac{\sqrt{2\eta+1}\sqrt{2\eta'+1}}{T}   \int_{0}^{\infty} \frac{J_{\eta+\frac{1}{2}}(2\pi \kappa) J_{\eta'+\frac{1}{2}}(2\pi \kappa)}{\kappa} \mathcal{T}\left( \frac{2}{T} \kappa \right) d\kappa
\end{equation}
and for $\eta = \eta'$, the variance in a single Legendre mode coefficient over the sample length $T$ is
\begin{equation}
\left\langle \left|a_\eta\right|^2 \right\rangle = \frac{2\eta+1}{T}  \int_0^{\infty} \frac{ J_{\eta+\frac{1}{2}}^2(2\pi \kappa)}{\kappa} \mathcal{T}\left( \frac{2}{T} \kappa \right) d\kappa.
\label{eqn:varcoeff}
\end{equation}

\bibliographystyle{apj}
\bibliography{specklelives}

\begin{thebibliography}{}
\expandafter\ifx\csname natexlab\endcsname\relax\def\natexlab#1{#1}\fi

\bibitem[{{Aime} {et~al.}(1986){Aime}, {Borgnino}, {Martin}, {Petrov}, \&
  {Ricort}}]{1986JOSAA...3.1001A}
{Aime}, C., {Borgnino}, J., {Martin}, F., {Petrov}, R., \& {Ricort}, G. 1986,
  Journal of the Optical Society of America A, 3, 1001

\bibitem[{{Beltramo-Martin} {et~al.}(2020){Beltramo-Martin}, {Ragland},
  {F{\'e}tick}, {Correia}, {Dupuy}, {Fiorentino}, {Fusco}, {Jolissaint},
  {Kamann}, {Marasco}, {Massari}, {Neichel}, {Schreiber}, \&
  {Wizinowich}}]{2020SPIE11448E..0AB}
{Beltramo-Martin}, O., {Ragland}, S., {F{\'e}tick}, R., {et~al.} 2020, in
  Society of Photo-Optical Instrumentation Engineers (SPIE) Conference Series,
  Vol. 11448, Society of Photo-Optical Instrumentation Engineers (SPIE)
  Conference Series, 114480A

\bibitem[{{Bloemhof} {et~al.}(2001){Bloemhof}, {Dekany}, {Troy}, \&
  {Oppenheimer}}]{2001ApJ...558L..71B}
{Bloemhof}, E.~E., {Dekany}, R.~G., {Troy}, M., \& {Oppenheimer}, B.~R. 2001,
  \apjl, 558, L71

\bibitem[{{Cavarroc} {et~al.}(2006){Cavarroc}, {Boccaletti}, {Baudoz}, {Fusco},
  \& {Rouan}}]{2006A&A...447..397C}
{Cavarroc}, C., {Boccaletti}, A., {Baudoz}, P., {Fusco}, T., \& {Rouan}, D.
  2006, \aap, 447, 397

\bibitem[{{Codona} \& {Kenworthy}(2013)}]{2013ApJ...767..100C}
{Codona}, J.~L., \& {Kenworthy}, M. 2013, \apj, 767, 100

\bibitem[{{Correia} {et~al.}(2017){Correia}, {Bond}, {Sauvage}, {Fusco},
  {Conan}, \& {Wizinowich}}]{2017JOSAA..34.1877C}
{Correia}, C.~M., {Bond}, C.~Z., {Sauvage}, J.-F., {et~al.} 2017, Journal of
  the Optical Society of America A, 34, 1877

\bibitem[{{Dainty} {et~al.}(1990){Dainty}, {Northcott}, \&
  {Qu}}]{1990JMOp...37.1247D}
{Dainty}, J.~C., {Northcott}, M.~J., \& {Qu}, D.~N. 1990, Journal of Modern
  Optics, 37, 1247

\bibitem[{{Fitzgerald} \& {Graham}(2006)}]{2006ApJ...637..541F}
{Fitzgerald}, M.~P., \& {Graham}, J.~R. 2006, \apj, 637, 541

\bibitem[{{Frazin}(2013)}]{2013ApJ...767...21F}
{Frazin}, R.~A. 2013, \apj, 767, 21

\bibitem[{{Frazin} \& {Rodack}(2021)}]{2021arXiv210506590F}
{Frazin}, R.~A., \& {Rodack}, A.~T. 2021, submitted to JATIS, arXiv:2105.06590

\bibitem[{{Galicher} {et~al.}(2010){Galicher}, {Baudoz}, {Rousset}, {Totems},
  \& {Mas}}]{2010A&A...509A..31G}
{Galicher}, R., {Baudoz}, P., {Rousset}, G., {Totems}, J., \& {Mas}, M. 2010,
  \aap, 509, A31

\bibitem[{{Goebel} {et~al.}(2018){Goebel}, {Guyon}, {Hall}, {Jovanovic},
  {Lozi}, \& {Martinache}}]{2018PASP..130j4502G}
{Goebel}, S.~B., {Guyon}, O., {Hall}, D. N.~B., {et~al.} 2018, \pasp, 130,
  104502

\bibitem[{{Gradshteyn} {et~al.}(2007){Gradshteyn}, {Ryzhik}, {Jeffrey}, \&
  {Zwillinger}}]{2007tisp.book.....G}
{Gradshteyn}, I.~S., {Ryzhik}, I.~M., {Jeffrey}, A., \& {Zwillinger}, D. 2007,
  {Table of Integrals, Series, and Products}

\bibitem[{{Guyon}(2005)}]{2005ApJ...629..592G}
{Guyon}, O. 2005, ApJ, 629, 592

\bibitem[{{Guyon}(2018)}]{2018ARAA..56..315G}
---. 2018, \araa, 56, 315

\bibitem[{{Haffert} {et~al.}(2021){Haffert}, {Males}, {Close}, {Van Gorkom},
  {Long}, {Hedglen}, {Guyon}, {Schatz}, {Kautz}, {Lumbres}, {Rodack}, {Knight},
  {Sun}, \& {Fogarty}}]{2021arXiv210307566H}
{Haffert}, S.~Y., {Males}, J.~R., {Close}, L.~M., {et~al.} 2021, arXiv
  e-prints, arXiv:2103.07566

\bibitem[{{Herscovici-Schiller} {et~al.}(2017){Herscovici-Schiller}, {Mugnier},
  \& {Sauvage}}]{2017MNRAS.467L.105H}
{Herscovici-Schiller}, O., {Mugnier}, L.~M., \& {Sauvage}, J.-F. 2017, \mnras,
  467, L105

\bibitem[{{Hinkley} {et~al.}(2007){Hinkley}, {Oppenheimer}, {Soummer},
  {Sivaramakrishnan}, {Roberts}, {Kuhn}, {Makidon}, {Perrin}, {Lloyd},
  {Kratter}, \& {Brenner}}]{2007ApJ...654..633H}
{Hinkley}, S., {Oppenheimer}, B.~R., {Soummer}, R., {et~al.} 2007, \apj, 654,
  633

\bibitem[{Kasdin(1995)}]{kasdin_falpha}
Kasdin, N. 1995, Proceedings of the IEEE, 83, 802

\bibitem[{{Lafreni{\`e}re} {et~al.}(2007){Lafreni{\`e}re}, {Marois}, {Doyon},
  {Nadeau}, \& {Artigau}}]{2007ApJ...660..770L}
{Lafreni{\`e}re}, D., {Marois}, C., {Doyon}, R., {Nadeau}, D., \& {Artigau},
  {\'E}. 2007, \apj, 660, 770

\bibitem[{{Macintosh} {et~al.}(2005){Macintosh}, {Poyneer}, {Sivaramakrishnan},
  \& {Marois}}]{2005SPIE.5903..170M}
{Macintosh}, B., {Poyneer}, L., {Sivaramakrishnan}, A., \& {Marois}, C. 2005,
  in Society of Photo-Optical Instrumentation Engineers (SPIE) Conference
  Series, Vol. 5903, Astronomical Adaptive Optics Systems and Applications II,
  ed. R.~K. {Tyson} \& M.~{Lloyd-Hart}, 170--177

\bibitem[{{Madec}(1999)}]{1999aoa..book.....R_ch6}
{Madec}, P.-Y. 1999, in Adaptive Optics in Astronomy, ed. F.~{Roddier}

\bibitem[{{Malbet} {et~al.}(1995){Malbet}, {Yu}, \&
  {Shao}}]{1995PASP..107..386M}
{Malbet}, F., {Yu}, J.~W., \& {Shao}, M. 1995, PASP, 107, 386

\bibitem[{{Males} \& {Guyon}(2018)}]{2018JATIS...4a9001M}
{Males}, J.~R., \& {Guyon}, O. 2018, Journal of Astronomical Telescopes,
  Instruments, and Systems, 4, 019001

\bibitem[{{Males} {et~al.}(2020){Males}, {Close}, {Guyon}, {Hedglen}, {Van
  Gorkom}, {Long}, {Kautz}, {Lumbres}, {Schatz}, {Rodack}, {Miller}, {Doelman},
  {Snik}, {Bos}, {Knight}, {Morzinski}, {Gasho}, {Keller}, {Haffert}, \&
  {Pearce}}]{2020SPIE11448E..4LM}
{Males}, J.~R., {Close}, L.~M., {Guyon}, O., {et~al.} 2020, in Society of
  Photo-Optical Instrumentation Engineers (SPIE) Conference Series, Vol. 11448,
  Society of Photo-Optical Instrumentation Engineers (SPIE) Conference Series,
  114484L

\bibitem[{{Marois} {et~al.}(2006){Marois}, {Lafreni{\`e}re}, {Doyon},
  {Macintosh}, \& {Nadeau}}]{2006ApJ...641..556M}
{Marois}, C., {Lafreni{\`e}re}, D., {Doyon}, R., {Macintosh}, B., \& {Nadeau},
  D. 2006, \apj, 641, 556

\bibitem[{{Marois} {et~al.}(2008){Marois}, {Lafreni{\`e}re}, {Macintosh}, \&
  {Doyon}}]{2008ApJ...673..647M}
{Marois}, C., {Lafreni{\`e}re}, D., {Macintosh}, B., \& {Doyon}, R. 2008, \apj,
  673, 647

\bibitem[{{Martinez} {et~al.}(2012){Martinez}, {Loose}, {Aller Carpentier}, \&
  {Kasper}}]{2012A&A...541A.136M}
{Martinez}, P., {Loose}, C., {Aller Carpentier}, E., \& {Kasper}, M. 2012,
  \aap, 541, A136

\bibitem[{{Milli} {et~al.}(2016){Milli}, {Banas}, {Mouillet}, {Mawet},
  {Girard}, {Vigan}, {Boccaletti}, {Kasper}, {Wahhaj}, {Lagrange}, {Beuzit},
  {Fusco}, {Sauvage}, \& {Galicher}}]{2016SPIE.9909E..4ZM}
{Milli}, J., {Banas}, T., {Mouillet}, D., {et~al.} 2016, in Society of
  Photo-Optical Instrumentation Engineers (SPIE) Conference Series, Vol. 9909,
  Adaptive Optics Systems V, ed. E.~{Marchetti}, L.~M. {Close}, \& J.-P.
  {V{\'e}ran}, 99094Z

\bibitem[{{Noll}(1976)}]{1976JOSA...66..207N}
{Noll}, R.~J. 1976, Journal of the Optical Society of America (1917-1983), 66,
  207

\bibitem[{Olver {et~al.}(2010)Olver, Lozier, Boisvert, \&
  Clark}]{Olver:2010:NHMF}
Olver, F.~W.~J., Lozier, D.~W., Boisvert, R.~F., \& Clark, C.~W., eds. 2010,
  {NIST Handbook of Mathematical Functions} (New York, NY: Cambridge University
  Press), print companion to \cite{NIST:DLMF}

\bibitem[{{Perrin} {et~al.}(2003){Perrin}, {Sivaramakrishnan}, {Makidon},
  {Oppenheimer}, \& {Graham}}]{2003ApJ...596..702P}
{Perrin}, M.~D., {Sivaramakrishnan}, A., {Makidon}, R.~B., {Oppenheimer},
  B.~R., \& {Graham}, J.~R. 2003, ApJ, 596, 702

\bibitem[{{Potier} {et~al.}(2020){Potier}, {Galicher}, {Baudoz}, {Huby},
  {Milli}, {Wahhaj}, {Boccaletti}, {Vigan}, {N'Diaye}, \&
  {Sauvage}}]{2020A&A...638A.117P}
{Potier}, A., {Galicher}, R., {Baudoz}, P., {et~al.} 2020, \aap, 638, A117

\bibitem[{{Poyneer} \& {Macintosh}(2006)}]{2006OExpr..14.7499P}
{Poyneer}, L.~A., \& {Macintosh}, B.~A. 2006, Optics Express, 14, 7499

\bibitem[{{Poyneer} {et~al.}(2007){Poyneer}, {Macintosh}, \&
  {V{\'e}ran}}]{2007JOSAA..24.2645P}
{Poyneer}, L.~A., {Macintosh}, B.~A., \& {V{\'e}ran}, J.-P. 2007, Journal of
  the Optical Society of America A, 24, 2645

\bibitem[{{Poyneer} {et~al.}(2016){Poyneer}, {Palmer}, {Macintosh},
  {Savransky}, {Sadakuni}, {Thomas}, {V{\'e}ran}, {Follette}, {Greenbaum},
  {Mark Ammons}, {Bailey}, {Bauman}, {Cardwell}, {Dillon}, {Gavel}, {Hartung},
  {Hibon}, {Perrin}, {Rantakyr{\"o}}, {Sivaramakrishnan}, \&
  {Wang}}]{2016ApOpt..55..323P}
{Poyneer}, L.~A., {Palmer}, D.~W., {Macintosh}, B., {et~al.} 2016, Appl.
  Optics, 55, 323

\bibitem[{{Racine} {et~al.}(1999){Racine}, {Walker}, {Nadeau}, {Doyon}, \&
  {Marois}}]{1999PASP..111..587R}
{Racine}, R., {Walker}, G. A.~H., {Nadeau}, D., {Doyon}, R., \& {Marois}, C.
  1999, \pasp, 111, 587

\bibitem[{{Rodack} {et~al.}(2021){Rodack}, {Frazin}, {Males}, \&
  {Guyon}}]{2021arXiv210506589R}
{Rodack}, A.~T., {Frazin}, R.~A., {Males}, J.~R., \& {Guyon}, O. 2021,
  submitted to JATIS, arXiv:2105.06589

\bibitem[{{Roddier} {et~al.}(1982){Roddier}, {Gilli}, \&
  {Lund}}]{1982JOpt...13..263R}
{Roddier}, F., {Gilli}, J.~M., \& {Lund}, G. 1982, Journal of Optics, 13, 263

\bibitem[{{Samland} {et~al.}(2021){Samland}, {Bouwman}, {Hogg}, {Brandner},
  {Henning}, \& {Janson}}]{2021A&A...646A..24S}
{Samland}, M., {Bouwman}, J., {Hogg}, D.~W., {et~al.} 2021, \aap, 646, A24

\bibitem[{{Scaddan} \& {Walker}(1978)}]{1978ApOpt..17.3779S}
{Scaddan}, R.~J., \& {Walker}, J.~G. 1978, \ao, 17, 3779

\bibitem[{{Seilmayer} {et~al.}(2020){Seilmayer}, {Garcia Gonzalez}, \&
  {Wondrak}}]{2020arXiv200110200S}
{Seilmayer}, M., {Garcia Gonzalez}, F., \& {Wondrak}, T. 2020, arXiv e-prints,
  arXiv:2001.10200

\bibitem[{{Singh} {et~al.}(2017){Singh}, {Lozi}, {Jovanovic}, {Guyon},
  {Baudoz}, {Martinache}, \& {Kudo}}]{2017PASP..129i5002S}
{Singh}, G., {Lozi}, J., {Jovanovic}, N., {et~al.} 2017, \pasp, 129, 095002

\bibitem[{{Soummer} {et~al.}(2007){Soummer}, {Ferrari}, {Aime}, \&
  {Jolissaint}}]{2007ApJ...669..642S}
{Soummer}, R., {Ferrari}, A., {Aime}, C., \& {Jolissaint}, L. 2007, \apj, 669,
  642

\bibitem[{{Soummer} {et~al.}(2012){Soummer}, {Pueyo}, \&
  {Larkin}}]{2012ApJ...755L..28S}
{Soummer}, R., {Pueyo}, L., \& {Larkin}, J. 2012, \apjl, 755, L28

\bibitem[{{Stangalini} {et~al.}(2017){Stangalini}, {Pedichini}, {Pinna},
  {Christou}, {Hill}, {Puglisi}, {Bailey}, {Centrone}, {Del Moro}, {Esposito},
  {Fiore}, {Giallongo}, {Hinz}, \& {Vaz}}]{2017JATIS...3b5001S}
{Stangalini}, M., {Pedichini}, F., {Pinna}, E., {et~al.} 2017, Journal of
  Astronomical Telescopes, Instruments, and Systems, 3, 025001

\bibitem[{{Steiger} {et~al.}(2021){Steiger}, {Currie}, {Brandt}, {Guyon},
  {Kuzuhara}, {Chilcote}, {Groff}, {Lozi}, {Walter}, {Fruitwala}, {Bailey},
  {Zobrist}, {Swimmer}, {Lipartito}, {Smith}, {Bockstiegel}, {Meeker},
  {Coiffard}, {Dodkins}, {Szypryt}, {Davis}, {Daal}, {Bumble}, {Vievard},
  {Sahoo}, {Deo}, {Jovanovic}, {Martinache}, {Tamura}, {Kasdin}, \&
  {Mazin}}]{2021arXiv210306898S}
{Steiger}, S., {Currie}, T., {Brandt}, T.~D., {et~al.} 2021, arXiv e-prints,
  arXiv:2103.06898

\bibitem[{{Veran} {et~al.}(1997){Veran}, {Rigaut}, {Maitre}, \&
  {Rouan}}]{1997JOSAA..14.3057V}
{Veran}, J.~P., {Rigaut}, F., {Maitre}, H., \& {Rouan}, D. 1997, Journal of the
  Optical Society of America A, 14, 3057

\bibitem[{{Walter} {et~al.}(2019){Walter}, {Bockstiegel}, {Brandt}, \&
  {Mazin}}]{2019PASP..131k4506W}
{Walter}, A.~B., {Bockstiegel}, C., {Brandt}, T.~D., \& {Mazin}, B.~A. 2019,
  \pasp, 131, 114506

\end{thebibliography}
\end{document}